\documentclass[graybox, envcountchap, authoryear, natbib]{svmult}

\usepackage{mathptmx}       
\usepackage{helvet}         
\usepackage{courier}        
\usepackage{type1cm}        
\usepackage{makeidx}         
\usepackage{graphicx}        
\usepackage{multicol}        
\usepackage[bottom]{footmisc}
\usepackage{enumerate}
\usepackage{amsmath}

\usepackage{natbib}

\newcommand {\hi} {{\rm H}\,{\small\rm I}}

\newcommand {\ha} {{\rm H}{\alpha}}

\newcommand {\siiii} {{\rm Si}\,{\small\rm III}}
\newcommand {\siiv} {{\rm Si}\,{\small\rm IV}}

\newcommand {\cii} {{\rm C}\,{\small\rm II}}
\newcommand {\ciii} {{\rm C}\,{\small\rm III}}
\newcommand {\civ} {{\rm C}\,{\small\rm IV}}

\newcommand {\kms} {\,{\rm km\,s}^{-1}}

\newcommand {\ergs} {\,{\rm erg\,s}^{-1}}
\newcommand {\erg} {\,{\rm erg}}

\newcommand {\cm} {\,{\rm cm}}

\newcommand {\cmsq}{\,{\rm cm^{-2}}}
\newcommand {\cmcu}{\,{\rm cm^{-3}}}

\newcommand {\K}{\,{\rm K}}

\newcommand {\pc} {\,{\rm pc}}
\newcommand {\kpc} {\,{\rm kpc}}

\newcommand {\mo}{\,M_{\odot}}

\newcommand {\kmskpc} {\,{\rm km\,s}^{-1}\,{\rm \kpc}^{-1}}

\newcommand {\yr}{\,{\rm yr}}
\newcommand {\Myr}{\,{\rm Myr}}
\newcommand {\Gyr}{\,{\rm Gyr}}

\newcommand {\moyr}{\,{{M}_\odot\,\rm yr}^{-1}}

\newcommand{\gtrsim}{\lower.7ex\hbox{$\;\stackrel{\textstyle>}{\sim}\;$}}
\newcommand{\gsim}{\lower.7ex\hbox{$\;\stackrel{\textstyle>}{\sim}\;$}}
\newcommand{\lsim}{\lower.7ex\hbox{$\;\stackrel{\textstyle<}{\sim}\;$}}
\newcommand{\lesssim}{\lower.7ex\hbox{$\;\stackrel{\textstyle<}{\sim}\;$}}
\newcommand{\grad}{\boldsymbol{\nabla}}

\newcommand {\apj}{\emph{ApJ} }
\newcommand {\aj}{\emph{AJ} }
\newcommand {\apjs}{\emph{AJSS} }
\newcommand {\apjl}{\emph{ApJL} }
\newcommand {\mnras}{\emph{MNRAS} }
\newcommand {\aap}{\emph{A\&A} }
\newcommand {\araa}{\emph{ARA\&A} }
\newcommand {\aapr}{\emph{A\&ARv} }
\newcommand {\pasp}{\emph{PASP} }

\newcommand {\nat}{\emph{Nature} }

\begin{document}

\title*{Gas Accretion via Condensation and Fountains} 
\author{Filippo Fraternali}
\institute{Filippo Fraternali \at University of Bologna, Department of Physics and Astronomy, via Berti Pichat 6/2, 40127, Bologna, Italy \email{filippo.fraternali@unibo.it}
\at University of Groningen, Kapteyn Astronomical Institute, Postbus 800, 9700 AV Groningen, The Netherlands}
%
%
\maketitle

\abstract{
For most of their lives, galaxies are surrounded by large and massive coronae of hot gas, which constitute vast reservoirs for gas accretion. 
This Chapter describes a mechanism that allows star-forming disc galaxies to extract gas from their coronae.
Stellar feedback powers a continuous circulation (galactic fountain) of gas from the disc into the halo, producing mixing between metal-rich disc material and metal-poor coronal gas.
This mixing causes a dramatic reduction of the cooling time of the corona making it condense and accrete onto the disc.
This \emph{fountain-driven accretion} model makes clear predictions for the kinematics of the extraplanar cold/warm gas in disc galaxies, which are in good agreement with a number of independent observations.
The amount of gas accretion predicted by the model is of the order of what is needed to sustain star formation. 
Accretion is expected to occur preferentially in the outer parts of discs and its efficiency drops for higher coronal temperatures. 
Thus galaxies are able to gather new gas as long as they do not become too massive nor fall into large halos and maintain their star-forming gaseous discs.
}

\section{Introduction}
\label{sec:intro}

Galaxies grow through the continuous acquisition of gas from the surrounding intergalactic environment.
This fundamental process of gas accretion is the key to understand galaxy formation and evolution and it has been described in all its most important aspects in the previous chapters of this book.
Here, we focus on a mechanism that appears to work well for the Milky Way and similar galaxies today but has the potential to be extended to other types of galaxies and epochs.
In this scheme, disc galaxies accrete gas from the surrounding massive hot corona, which is expected to extend out to the virial radius \citep{Fukugita&Peebles2004}.
The cooling and accretion occurs very close to the galactic disc and it is triggered and regulated by stellar feedback.
Very strong and efficient feedback, like in starburst galaxies where we observe strong galactic winds, may have the effect of inhibiting accretion and heating the corona \citep{Sokolowska+2016}.
However, more common low-levels of feedback, 
typical of a galactic fountain regime in main-sequence galaxies, has the opposite effect of stimulating gas accretion.
This overlooked effect of supernova feedback is the subject of this Chapter.

Stellar winds and supernovae produce large expanding bubbles that bring large amounts of material from the disc into the halo \citep{MacLow+1989}.
These phenomena set up a circulation \citep[galactic fountain][]{Shapiro&Field1976, Fraternali&Binney2006} that continuously lifts disc gas into the halo causing it to mix with the hot ($T\gtrsim 10^6 \K$) gas of the corona.
The disc gas is more metal-rich and colder than the hot coronal gas and their mixing dramatically reduces the cooling time of the hot gas leading to the condensation and accretion of a significant fraction of the lower corona.
This mechanism leaves a clear signature (lagging rotation) on the kinematics of the extraplanar cold/warm ($T\lesssim 10^{5} \K$) gas in disc galaxies, which is observed in local galaxies.
The modelling of these data leads to the estimate of the amount of gas accretion via condensation and fountains. 
Remarkably, this turns out to be of the order of the star formation rate (SFR), providing a viable solution of the long-standing problem of the feeding of star formation in disc galaxies.
In the literature, this mechanism is referred to as \emph{supernova-driven accretion} or fountain-driven accretion, here we simply call it \emph{fountain accretion}. 
Before embarking on the description of the physical phenomenon and its observational evidence, we review the main lines of thought that brought about the formulation of this theory and the key ingredients that are required to make it work.

A number of lessons can be indirectly learned about gas accretion by studying the stellar and gaseous components of galaxies and their star formation \citep{SanchezAlmeida+2014}; we briefly summarize them here.
\begin{enumerate}[1.]
\item{Star-forming galaxies have SFRs that are typically not sustainable in isolation for a Hubble time, i.e.\ the ratio between their gas mass and their SFR (\emph{depletion time}) is of the order of a Gigayear for every galaxy at every redshift \citep[e.g.][]{Saintonge+2013}. This clearly points to the need of continuous gas accretion \citep{Fraternali&Tomassetti2012}.
}
\item{Galaxies, especially when going through starburst phases, have strong winds that can remove gas \citep{Veilleux+2005} making their need for accretion even more severe.
}
\item{Galaxy discs grow inside-out, even today \citep{Pezzulli+2015}, in the sense that the size of their stellar disc increases with time. 
This points to late accretion of gas preferentially in the outer parts and therefore at high angular momentum \citep{Goldbaum+2015}.
}
\item{Chemical evolution models require late accretion of low-metallicity gas, typically at or below 0.1 Solar \citep{Tosi1988b} in order to explain the chemistry of stars in the disc of the Milky Way \citep{Chiappini+2001,Schoenrich&Binney2009, Kubryk+2013}.
}
\item{Galaxies progressively evolve from the blue cloud to the red sequence with no evidence of significant going back \citep[e.g.][]{Peng+2010}. 
About $80\%$ of red-sequence galaxies are not detected in deep $\hi$ surveys \citep{Catinella+2012}.
}
\item{Finally, galaxies have much less (factor 5 or more) baryonic matter in the form of cold gas and stars than we would expect given their dark-matter masses \citep{Papastergis+2012,vanUitert+2016}.
These \emph{missing baryons} are at least partially located in the hot metal-poor coronae and thus being hard to detected \citep{Bregman2007}.
Only rich galaxy clusters contain the expected amount of baryons mostly in the form of hot gas \citep[e.g.][]{McGaugh+2010}.
}
\end{enumerate}

The picture that emerges from above is that as long as they keep forming stars at the observed levels, disc galaxies must continuously gather low-metallicity and relatively high-angular momentum gas from the surrounding medium.
This gas accretion proceeds until some form of star-formation quenching occurs or the gas supply is somehow cut off \citep[e.g.][]{Lilly+2013}. 
This phenomenon is clearly more efficient in dense environments where the causes for gas removal are better understood, e.g.\ ram pressure stripping and tidal interactions \citep[e.g.][]{Fasano+2000}. 
After quenching either the gas reservoir is not available anymore or the galaxy loses the ability to harvest material from its reservoir and thereafter evolves passively as a red-sequence galaxy.

The above indirect evidence must be combined with direct observations of gas around galaxies to get a consistent picture of the census of baryons around galaxies and their physical state.
These observations have been described in detail in the first part of this book, here we summarize those most relevant for fountain accretion.
\begin{enumerate}[a.]
\item{Cold ($T\lesssim 10^4 \K$) and dense gas (visible in $\hi$ emission) is only observed in large quantities very close (few kpc) to galaxy discs.
This is the so-called \emph{extraplanar gas} that we describe in some detail in \S \ref{sec:epg}.
At larger distances, some $\hi$ clouds are observed that could be part of the same medium (see \S \ref{sec:observations}).
In the Milky Way, they are called \emph{High-Velocity Clouds} \citep[HVCs][]{Wakker&vanWoerden1997} and typically extend to $\sim 10\kpc$ from the Galactic disc.
The Galactic HVCs have relatively low metallicity \citep{vanWoerden+2004} and appear to be falling onto the disc, however their masses are typically $1-2$ orders of magnitude smaller than what would be needed to feed the star formation \citep{Putman+2012}.
Further out, there is no evidence of large ``free floating'' $\hi$-emitting clouds down to masses of $M_{\rm \hi}\sim 10^6 \mo$ \citep[e.g.][]{Pisano+2011}.
}
\item{Large masses of neutral gas away from galaxies are only observed in the presence of strong interactions (encounters/mergers) or in tight groups \citep{Yun+1994, Putman+2003, Lewis+2013}.
Here we do not discuss these sporadic phenomena, which may also have relevance in promoting gas accretion from one galaxy to another \citep{Nichols+2011, Fox+2014}.
Estimates of neutral gas accretion from minor mergers today return values that are orders of magnitude below the requirements to sustain star formation \citep{DiTeodoro&Fraternali2014}.
Moreover, the nearly constant amount of $\hi$ in galaxies across the whole Hubble time \citep[e.g.][]{Zafar+2013} strongly suggests that most gas accretion must come from a gas phase different from $\hi$.
}
\item{Warm/ionized ($10^4< T< 10^6 \K$) gas is ubiquitously detected both around the Milky Way \citep{Sembach+2003,Shull+2009,Lehner+2012, Richter+2013} and around external galaxies \citep{Werk+2013,Werk+2014} via absorption towards distant quasars.
This component has too low temperature to be in hydrostatic equilibrium in the potential of Milky Way-like galaxies, thus it is likely infalling or outflowing.
The mass contained in this component is hard to estimate but it could provide accretion rates of the order of $\sim 1 \moyr$ \citep{Lehner&Howk2011}. 
}
\item{HVCs both neutral and ionized must be confined by a hot medium generally referred to as the \emph{galactic corona}.
Recent observations in the X-rays have revealed the presence of this hot corona both around the Milky Way \citep{Miller&Bregman2015} and around external massive spiral galaxies \citep{Dai+2012,Bogdan+2013}.
The corona should extend out to the virial radius of the dark matter halo and contain a mass at least as large as the baryonic mass in the galaxy disc \citep{Gatto+2013}.
}
\end{enumerate}

In what follows we try to incorporate most of the direct and indirect evidence into a coherent picture. 
We first give a general description of the properties of the extraplanar gas (\S \ref{sec:epg}), the explanation of which generates problems for standard galactic fountain models pointing to the importance of interaction between disc and coronal gas at the disc-halo interface (\S \ref{sec:fountain}).
We describe the hydrodynamical simulations of disc-corona mixing that predict the condensation of the coronal gas (\S \ref{sec:simulations}) and this leads us to the description of the fountain-driven accretion model (\S \ref{sec:fountainAccretion}).
We then give a full account of the comparison of the predictions of this model with the observations (\S \ref{sec:observations}) and finally, we discuss the potential relevance of this model for galaxy evolution focusing in particular on the onset of metallicity gradients and star-formation quenching (\S \ref{sec:evolution}).

\section{Extraplanar gas: life at the disc-halo interface}
\label{sec:epg}

The discs of nearby spiral galaxies with sufficient star formation are surrounded by thick layers of neutral and ionized gas that extend kiloparsecs from the plane \citep{Fraternali+2002, Matthews&Wood2003, Oosterloo+2007, Zschaechner+2015}.
This layer is also observed in the Milky Way \citep{Lockman2002, Marasco&Fraternali2011}. 
The $\hi$ in the extraplanar gas is dense enough to be observed in emission in deep observations \citep{Sancisi+2008} as is the ionized gas \citep{Rossa&Dettmar2003, Heald+2007, Kamphuis2008}.
This is of fundamental importance for what follows because these observations can be used to study the detailed kinematics of the gas at the disc-halo interface and understand its origin. 

\begin{figure}[ht]
\sidecaption
\includegraphics[trim=0cm 0cm 0cm 0cm, clip=true, angle=-90, width=\textwidth]{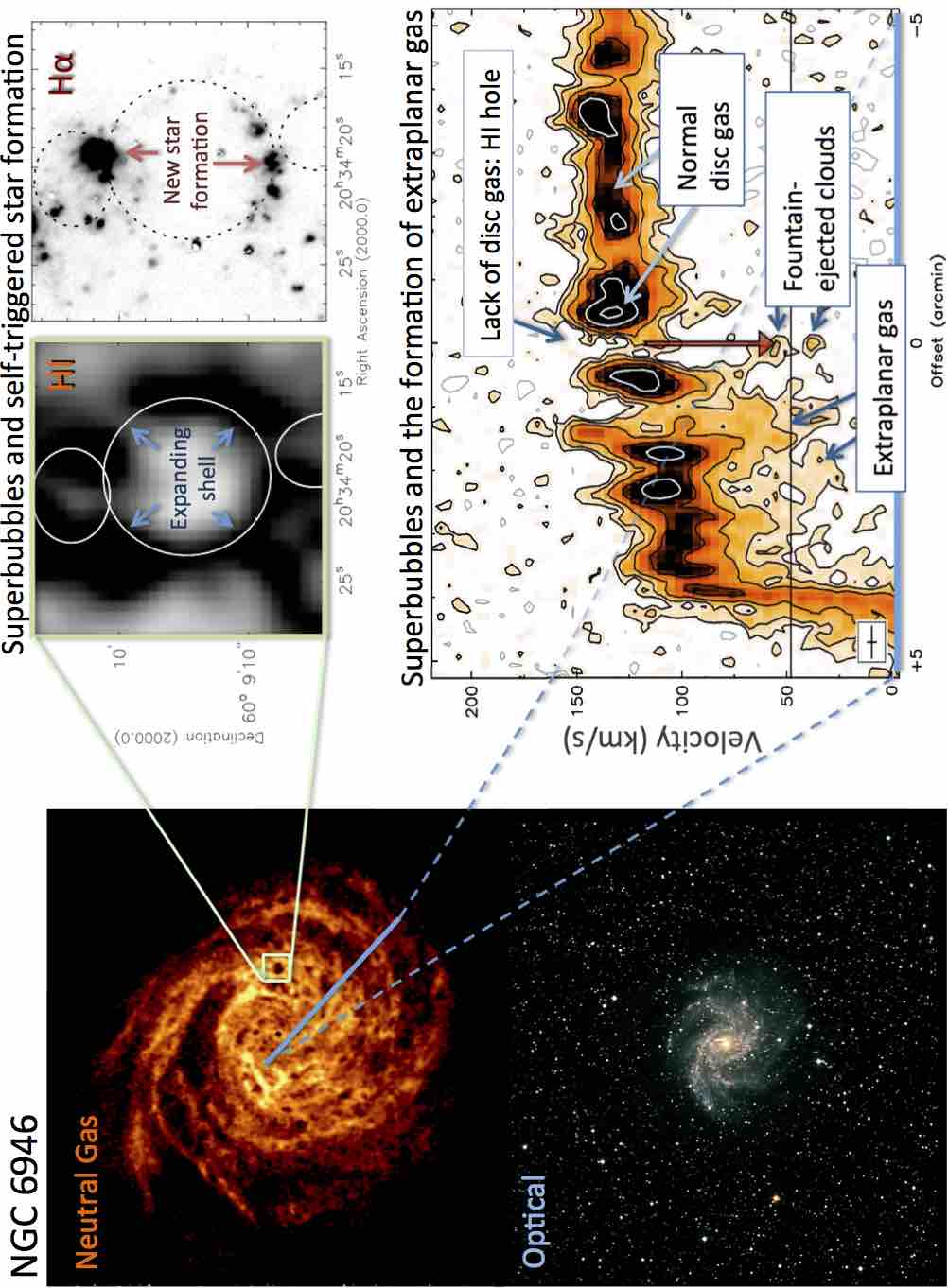}
\caption{The formation of the extraplanar gas layer from superbubble blowout in the nearby disc galaxy NGC\,6946.
Left panels: comparison between the extent of the $\hi$ disc in this galaxy and the bright optical disc \citep[adapted from][]{Boomsma+2008}.
Top right: the expansion of a large superbubble has produced a hole in the $\hi$ distribution in the disc and caused the compression of the surrounding gas triggering a new generation of stars, testified by the intense $\ha$ emission observed at the edges of the hole.
Bottom right: position-velocity plot along the line shown overlaid on the top-left total $\hi$ map.
In regions of depressed emission in the disc ($\hi$ holes) fast-moving clouds are observed leaving the disc and joining the diffuse material that makes up the extraplanar gas layer.
Note how the extraplanar gas is located at velocities between the normal disc gas and the systemic velocity of the galaxy (horizontal line), i.e.\ it rotates more slowly that the disc gas.
}
\label{fig:ngc6946}
\end{figure}

The extraplanar gas in disc galaxies has three main properties: i) it is located considerably above the standard scale-height expected for the $\hi$ layer (of order $100 \pc$) and makes up about $10-30\%$ of the total $\hi$ of a galaxy;
ii) it shows a distinctive kinematics with respect to the disc gas, in particular it rotates more slowly \citep{Fraternali+2005} and it shows non circular motions \citep{Fraternali+2001};
iii) it is located mostly in the star-forming inner part of the galaxies.
The first detection of a widespread layer of extraplanar gas with coherent kinematics (low rotation) was reported for the edge-on galaxy NGC\,891 \citep{Swaters+1997}.
Then, thanks to its peculiar kinematics it has been detected and studied in galaxies not seen edge-on, in particular in NGC\,2403 \citep{Schaap+2000, Fraternali+2001}.
In these intermediate-inclination galaxies, it shows up as a smooth emission at lower rotation velocities with respect to the disc emission, in particular along the major axis of the galaxy (see position-velocity plot in Fig.\ \ref{fig:ngc6946}).
The peculiar kinematics of the extraplanar gas also allowed it to be identified in the Milky Way \citep{Marasco&Fraternali2011}.

As we explain in \S \ref{sec:fountain}, the origin of a significant fraction ($\sim 80\%$) of the extraplanar gas is likely to be stellar feedback.
The combined action of stellar winds and supernovae in large star clusters (OB associations) produces the formation and expansion of large bubbles called \emph{superbubbles} \citep{MacLow&McCray1988}.
Unlike normal wind/supernova bubbles, superbubbles can easily reach sizes that exceed the typical thickness of the disc of cold gas.
When the size of a superbubble shell (supershell) becomes of the order of the disc thickness, the expansion accelerates vertically and a blowout occurs \citep{MacLow+1989}.
The blowout has two important consequences.
First, it leaves \emph{holes} in the distribution of the cold neutral gas.
A large number of these $\hi$ holes are observed in nearby nearly face-on galaxies \citep{Puche+1992, Boomsma+2008}.
Second, it ejects gas from the disc into the halo \citep{Norman&Ikeuchi1989}. 
This gas is both \emph{cold} ($T\sim 10^4 \K$), coming from the supershell and \emph{hot} ($T\gtrsim 10^6 \K$) from the bubble interior.
Figure \ref{fig:ngc6946} illustrates these phenomena in one galaxy, NGC\,6946, seen at the relatively low inclination of 30 degrees with respect to the line of sight.
$\hi$ clouds ejected from the disc of this galaxy, often at the locations of $\hi$ holes are sustaining a layer of extraplanar neutral gas.

A somewhat more obvious way to visualize the extraplanar gas is to observe edge-on galaxies.
Figure \ref{fig:ngc891} shows deep $\hi$ observations (blue shade and contours) of the edge-on spiral NGC\,891 overlaid on an optical image \citep{Oosterloo+2007}.
The $\hi$ layer extends above the plane to remarkable distances with a scale-height $h\simeq 2.2 \kpc$.
The total mass of $\hi$ located outside the disc is about $1.2\times 10^9 \mo$, which corresponds to almost $30\%$ of the total $\hi$ mass.
This galaxy is likely to be on the high end of the spectrum of extraplanar gas properties, in particular for its large extraplanar $\hi$ mass.
Preliminary results obtained by the HALOGAS survey \citep{Heald+2011} show several other galaxies with relatively high SFR have typical extraplanar $\hi$ masses of order $10-20\%$ of their total $\hi$ masses (Fraternali et al., in prep.).

\section{Galactic fountains and the origin of extraplanar gas}
\label{sec:fountain}

Since its discovery, 
three models have been proposed to explain the cold extraplanar gas: hydrostatic equilibrium, accretion/inflow from the circumgalactic medium and galactic fountain.
The hypothesis that the extraplanar gas could be in equilibrium has been investigated with two different approaches, to which we can refer as hydrodynamical and dynamical.
In the hydrodynamical model the gas layer is a baroclinic fluid\footnote{We recall that this is a fluid with an equation of state where pressure does not depend on density alone. This is in contrast to a barotropic fluid where $P=P(\rho)$.} in hydrostatic equilibrium with the galactic potential \citep{Barnabe+2006}.
Despite the success in reproducing the extraplanar kinematics, this model predicts temperatures for the gas larger than $10^5 \K$, which are incompatible with the neutral gas phase.
The second attempt assumed that the extraplanar layer consists of a large number of gas clouds characterized by large relative velocities, i.e.\ a velocity dispersion much higher than that of the disc $\hi$.
The equilibrium of such a medium was studied using the Jeans equations and their predictions compared in the 3D space (position on the sky and line-of-sight velocity) to the observed $\hi$ emission of NGC\,891 \citep{Marinacci+2010}.
It was found that the data are incompatible with a system of clouds with an isotropic velocity dispersion and the only configuration that resembles the observations is one with a strong vertical anisotropy ($\sigma_z\gg\sigma_{\phi}\sim\sigma_{R}$).
This is akin to the anisotropy that one finds in galactic fountain models.

The possibility that the extraplanar gas is produced exclusively by extragalactic inflow has been studied using hydrodynamical simulations of galaxy formation \citep{Kaufmann+2006}.
A model of this kind reproduces the kinematics of the extraplanar gas, in particular the rotation curves of NGC\,891 \citep{Fraternali+2005}.
This model is however unrealistic for a number of reasons.
First, the cold clouds that formed in these early SPH simulations are now known to be numerical artifacts and not the product of thermal instabilities as initially thought \citep{Kaufmann+2009, Read&Hayfield2012}.
In fact, the classical Field instability should not be at work at all in the coronae of these galaxies because it is counteracted by buoyancy and suppressed by thermal conduction \citep{Binney+2009, Joung+2012}.
Moreover, pure infall models would give totally unrealistic infall rates, for example of the order of $30\moyr$ in NGC\,891, an order of magnitude larger than the galaxy's SFR.

\begin{figure}[ht]
\sidecaption
\includegraphics[trim=0cm 0cm 0cm 0cm, clip=true, angle=-90, width=\textwidth]{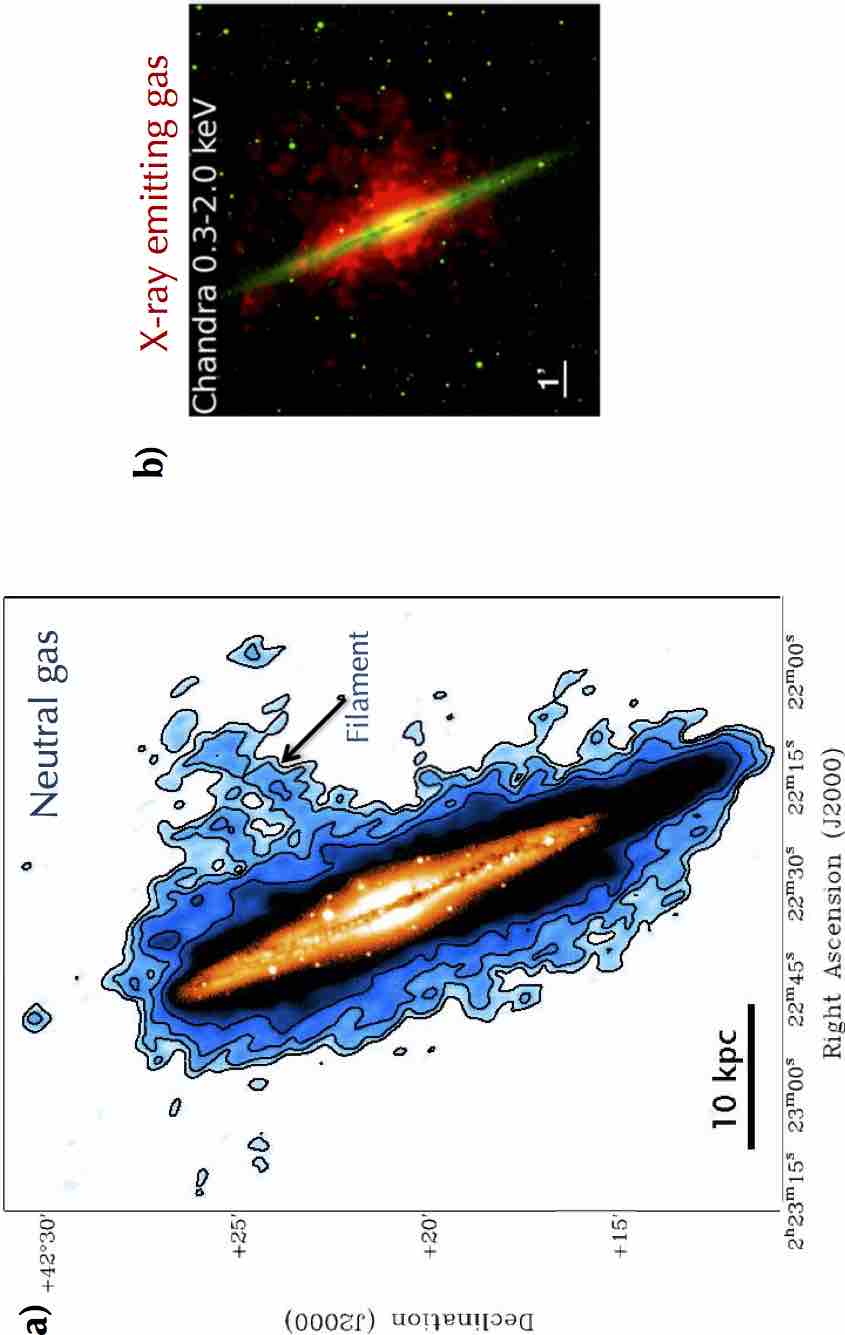}
\caption{Comparison between cold and hot gas in the halo region close to the disc of the edge-on galaxy NGC\,891, the two images are on the same scale.
\emph{a)} total $\hi$ map from the deep observations obtained with the Westerbork Synthesis Radio Telescope (WSRT) \citep{Oosterloo+2007} overlaid on an optical image (orange).
The neutral cold ($T<2\times10^4 \K$) gas extends up to $\sim 8 \kpc$ everywhere above and below the inner disc with a feature (filament) reaching up to $20 \kpc$.
The total $\hi$ mass of extraplanar gas is of about $1.2\times 10^9\mo$ ($1.6\times 10^9\mo$ taking helium into account).
\emph{b)} Chandra X-ray observations of NGC\,891 overlaid on an optical image \citep[from][]{Hodges-Kluck&Bregman2013}.
The X-ray emitting hot ($T=4\times10^6 \K$) gas appears more centrally concentrated and rounder in shape than the $\hi$ emission.
Its total mass (assuming a cylindrical halo with scale-height $5\kpc$) is $1-3 \times 10^8 \mo$.
Note that this is only the very inner part of the galactic corona, which likely extends, albeit with much lower densities (and therefore undetectable with these observations), out to the virial radius (see text).}
\label{fig:ngc891}
\end{figure}

The third possibility that we can consider is that the extraplanar gas is the product of a widespread galactic fountain.
As mentioned in \S \ref{sec:epg}, the blowout of superbubbles is expected to eject a large mass of gas out of the plane of the disc. 
The lack of this gas in the disc is observed in the form of $\hi$ holes and, in fact, the mass \emph{missing} from the holes is comparable to the mass of the extraplanar gas layer \citep{Boomsma+2008}.
In the first models of galactic fountain, the general idea was that the gas is ejected hot from the disc and then it eventually cools down to produce a \emph{rain} of $\hi$ clouds \citep{Shapiro&Field1976, Bregman1980}.
However, current observations show that most of the gas in the region close to the galactic disc is in fact at relatively low temperatures.
The example of NGC\,891 is very instructive.
Figure \ref{fig:ngc891} shows the total $\hi$ map of this galaxy compared with the deepest X-ray observations with Chandra \citep{Hodges-Kluck&Bregman2013}, on the same scale.
Clearly, the neutral gas appears more widespread and even more extended in the vertical direction than the hot X-ray emitting gas.
The mass of neutral extraplanar gas is $M_{\rm cold}\simeq 1.6\times 10^9\mo$ \citep{Oosterloo+2007} including the correction for helium abundance.
In roughly the same region, the mass of hot gas is instead at most a few times $10^8\mo$ (see Table \ref{tab:epg}), about an order of magnitude lower.
This simple evidence shows that most of the circulation of gas in the extraplanar layer occurs at relatively low temperatures, either because the gas is ejected cold \citep{Fraternali2012, Hodges-Kluck&Bregman2013} or because it cools rapidly \citep{Houck&Bregman1990}.
The same is true for other galaxies in which extraplanar gas has been studied in detail.
In table \ref{tab:epg} we report the properties of four representative galaxies.
A comprehensive compilation of the properties of the extraplanar gas in galaxies is currently lacking, only partial lists exist \citep{Sancisi+2008,Fraternali2009,Zschaechner+2015}.

\begin{table}
\caption{Property of the extraplanar gas in four representative galaxies}
\label{tab:epg}
\begin{center}
\begin{tabular}{lcccccccc}
\hline\noalign{\smallskip}
Galaxy & $M_{\hi}$ & $\frac{M_{\hi}}{M_{\hi,\,{\rm disc}}}$ & Gradient & $M_{\rm warm}$ & $T_{\rm Xray}$ & $M_{\rm Xray}^{~~~~~~a}$ & $Z_{\rm Xray}$ & Ref. \\
 & ($10^8\mo$) & ($\%$) & ($\kmskpc$) & ($10^8\mo$) & ($10^6\K$) & ($10^8\mo$) & ($Z_{\odot}$) \\
\noalign{\smallskip}\svhline\noalign{\smallskip}
NGC\,891 & $12$ & $30\%$  & 15 & few & $3-4$ & $1-3$ & 0.1 & (a, b, c)\\ 
Milky Way & $3^b$ & $10\%$ & 15 & 1.1 & $3$ & --$~^d$ & $0.3$ & (d, e, f) \\    
NGC\,2403 & $3^b$ & $10\%$  & $10-15$ & yes & $2-8$ & $\sim 0.1 $ & -- & (g, h, i) \\
NGC\,6946 & $3-10^b$ & $4-15\%$  & -- & yes & 4 & $0.4-0.8^c$ & -- & (j, l) \\    
\noalign{\smallskip}\hline\noalign{\smallskip}
\end{tabular}\\
\end{center}
$^a$ This is only the mass of the X-ray emitting gas in the region close to the disc, which roughly overlaps with the region of the extraplanar neutral gas. $^b$ These masses may be underestimated because some extraplanar gas emission overlaps with the disc gas in these datacubes.
$^c$ Assuming a range of scale-heights between $1$ and $5\kpc$ \citep[see][]{Schlegel+2003}.
$d$ The density profile used for the hot gas in \citet{Miller&Bregman2015} does not allow a reliable estimate of the inner mass.
References: (a) \citet{Oosterloo+2007}, (b) \cite{Howk&Savage2000}, (c) \cite{Hodges-Kluck&Bregman2013},
(d) \cite{Marasco&Fraternali2011}, (e) \cite{Lehner&Howk2011}, (f) \cite{Miller&Bregman2015},
(g) \cite{Fraternali+2002}, (h) \cite{Fraternali+2004}, (i) \cite{Fraternali+2002},
(j) \cite{Boomsma+2008}, (l) \cite{Schlegel+2003}.
\end{table}

The most straightforward model of a galactic fountain is the so-called \emph{ballistic model} \citep{Collins+2002, Spitoni+2008}.
In this formulation, fountain clouds are treated like test particles shot up nearly vertically from the disc by stellar feedback. 
They follow trajectories governed only by gravity and return to another location in the disc where they join the local interstellar medium.
This treatment is clearly limited as it neglects the hydrodynamical nature of the fountain gas but it is a very useful base model to study the physical phenomenon \citep{Fraternali&Binney2006}.

\begin{figure}[ht]
\begin{center}
\includegraphics[trim=0cm 0cm 0cm 0cm, clip=true, angle=0, width=0.52\textwidth]{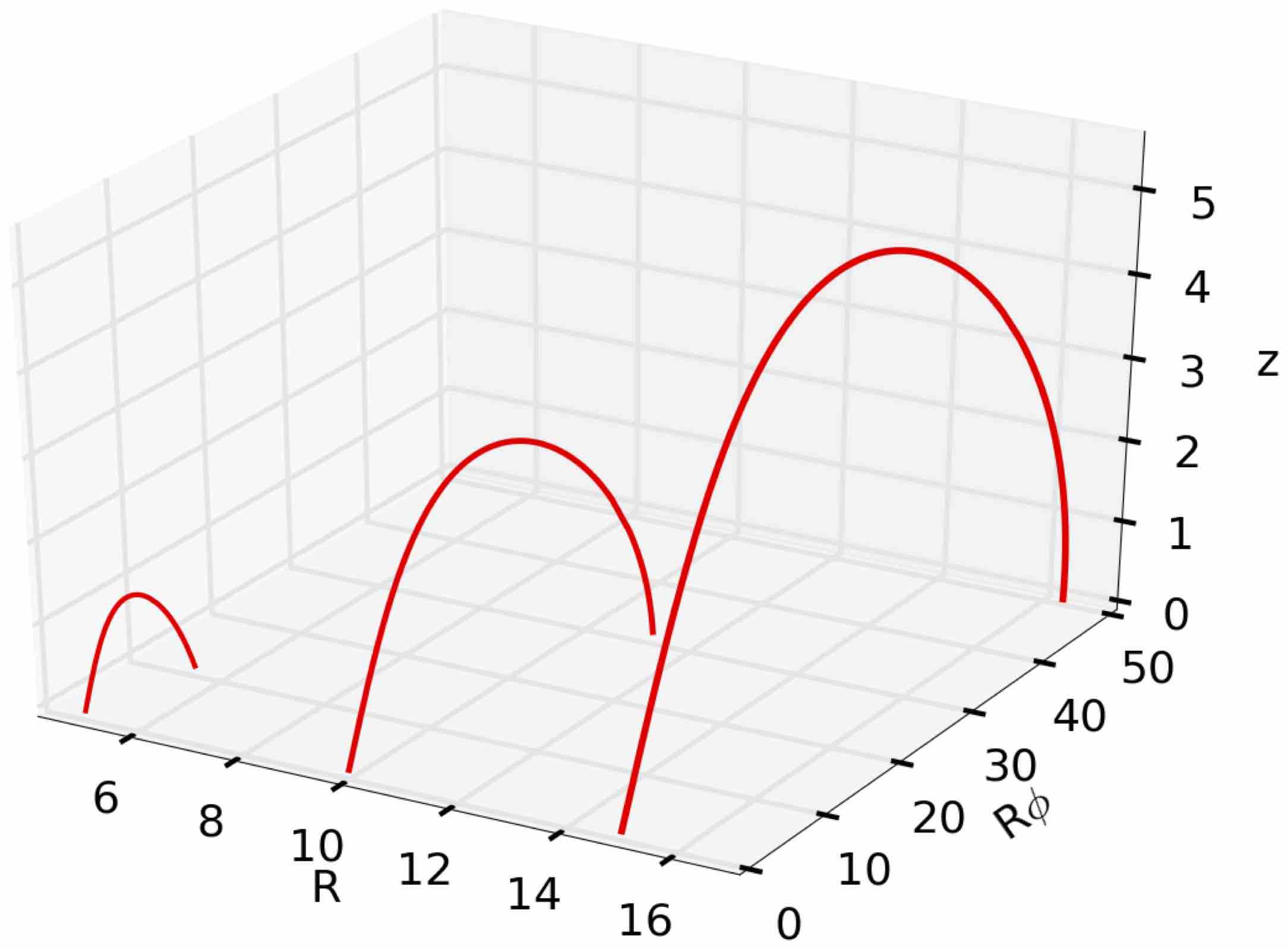}
\includegraphics[trim=0cm 0cm 0cm 0cm, clip=true, angle=0, width=0.47\textwidth]{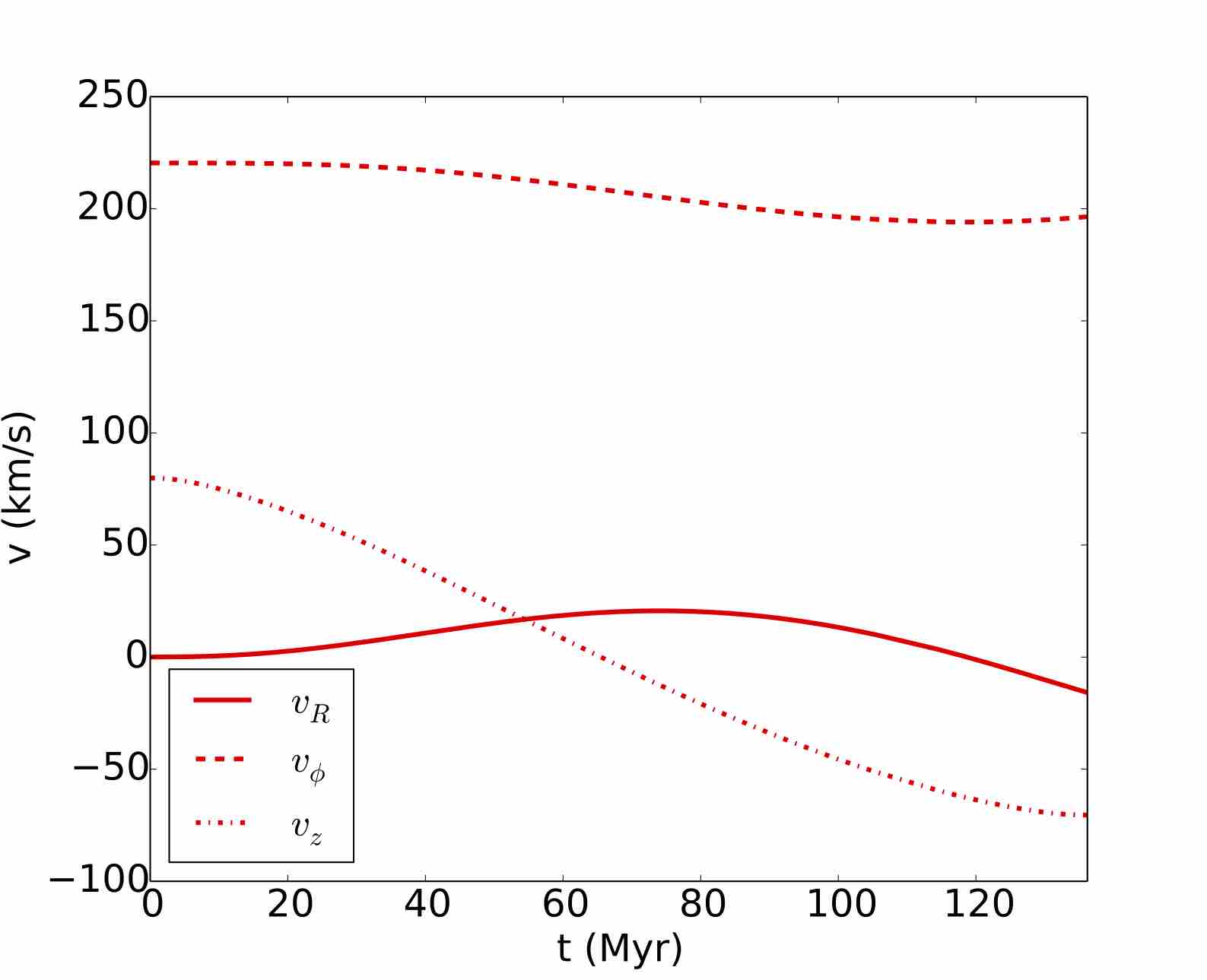}
\caption{
\emph{Left:} Orbits of galactic fountain particles in the potential of the Milky Way shot vertically upwards with $v_{\rm kick}=75\kms$ from three radii $R=5,\,10,\,15 \kpc$.
All particles fall back to the disc at slightly larger radii with respect to the kick radii. 
Note how the maximum height reached varies with radius due to the decrease of the vertical force.
Only gravity is included in this calculation.
\emph{Right:} Cylindrical components of the velocity of the fountain particle ejected from $R=10 \kpc$.
}
\end{center}
\label{fig:orbitsPure}
\end{figure}

Once the galactic potential is fixed, the most important parameter of the model is the \emph{kick velocity} of the fountain particles.
Luckily, some of the observable properties strongly depend on the kick velocity and so its value can be accurately constrained.
First, as one may expect, the higher the kick velocity the larger the scale-height of the extraplanar layer produced by the fountain.
This led to an estimate of the average values of the kick velocity of about $v_{\rm kick}=70-80\kms$ for NGC\,891 \citep{Fraternali&Binney2006} and the Milky Way \citep{Marasco+2012}.
These values are fully compatible with the velocities of the material in the shell of a superbubble at blowout coming from hydrodynamical simulations \citep{MacLow+1989, Melioli+2008}.
In practice, in a galactic fountain model, clouds are ejected at different radii with a characteristic velocity $v_{\rm kick}$ that is usually kept constant with radius. 
Figure \ref{fig:orbitsPure} (left) shows the orbits that one obtains by shooting fountain particles vertically straight up in the potential of the Milky Way starting from three representative radii.
Generally, the particles tend to move outward.
This happens because, as a consequence of the kick, the particle acquires a velocity that always exceeds its circular velocity $\mathbf{v}=\mathbf{v}_{\rm c}+\mathbf{v}_{\rm kick}>\mathbf{v}_{\rm c}$ and thus it finds itself at the ``pericenter'' of the orbit (in the approximation of a nearly spherical potential).
The maximum height reached varies substantially with radius reaching distances of up to several kpc if the ejection occurs at large radii.

Figure \ref{fig:orbitsPure} (right) shows the three cylindrical components of the velocity of the particle ejected at $R=10 \kpc$.
As one may expect for ballistic trajectories, $v_{\rm z}$ has a sinusoidal shape with the particle falling back to the disc at a speed roughly equal to the ejection velocity. 
The radial component $v_{\rm R}$ is mostly positive, given that the particle moves outward. 
Of great interest is the azimuthal component $v_{\phi}$, which decreases slightly.
This decrease is explained by angular momentum conservation.
It can be shown \citep{Marinacci+2010} that the value of $v_{\phi}$ at each location along the orbit is similar to the ``circular velocity'' at that location:
\begin{equation}
v_{\rm c}(R,z)=\left[R\frac{\partial \Phi(R,z')}{\partial R}\right]^{1/2}_{(z'=z)}.
\label{eq:vc}
\end{equation}
where $\Phi$ is the axisymmetric galactic potential and the derivative is calculated at height $z$.
This velocity provides a reference for the comparison between the prediction of the models and the rotation of the extraplanar gas.

The global model of a galactic fountain is constructed by integrating a large number of orbits like the ones described above.
The first versions of the model were fully axisymmetric (but see end of \S \ref{sec:observations}). 
The clouds are shot with a Gaussian distribution of velocities with dispersion $v_{\rm kick}$ and a relatively small opening angle \citep[estimated directly from observations;][]{Fraternali&Binney2006} with respect to the normal direction.
The number of clouds ejected at each radius is proportional to the star formation rate surface density or, equivalently, the supernova rate density \citep{Fraternali&Binney2006}. 
Once a suitable number of orbits have been integrated in the meridional plane ($R,z$), they are randomly distributed in azimuth.
The 21-cm emission from each of these ``clouds'' is generated by assuming that they are partially or totally neutral at each time step along their orbits.
A fraction of the material escaping from the superbubbles is likely to be photoionized given the strong radiation field -- see for instance the Ophiucus superbubble in the Milky Way \citep{Pidopryhora+2007}.
To take this into account one introduces a second parameter: the \emph{ionized fraction} ($f_{\rm ion}$) of the ascending fountain clouds. The descending clouds are usually considered neutral.
If $f_{\rm ion}=0$ the ascending clouds are always visible in $\hi$, otherwise a fraction of their orbit will be invisible.
Both $v_{\rm kick}$ and $f_{\rm ion}$ are parameters that can be fitted to observations (\S \ref{sec:observations}, Table \ref{tab:evidence}).

\begin{figure}[ht]
\sidecaption
\includegraphics[trim=0cm 0cm 0cm 0cm, clip=true, angle=-90, width=\textwidth]{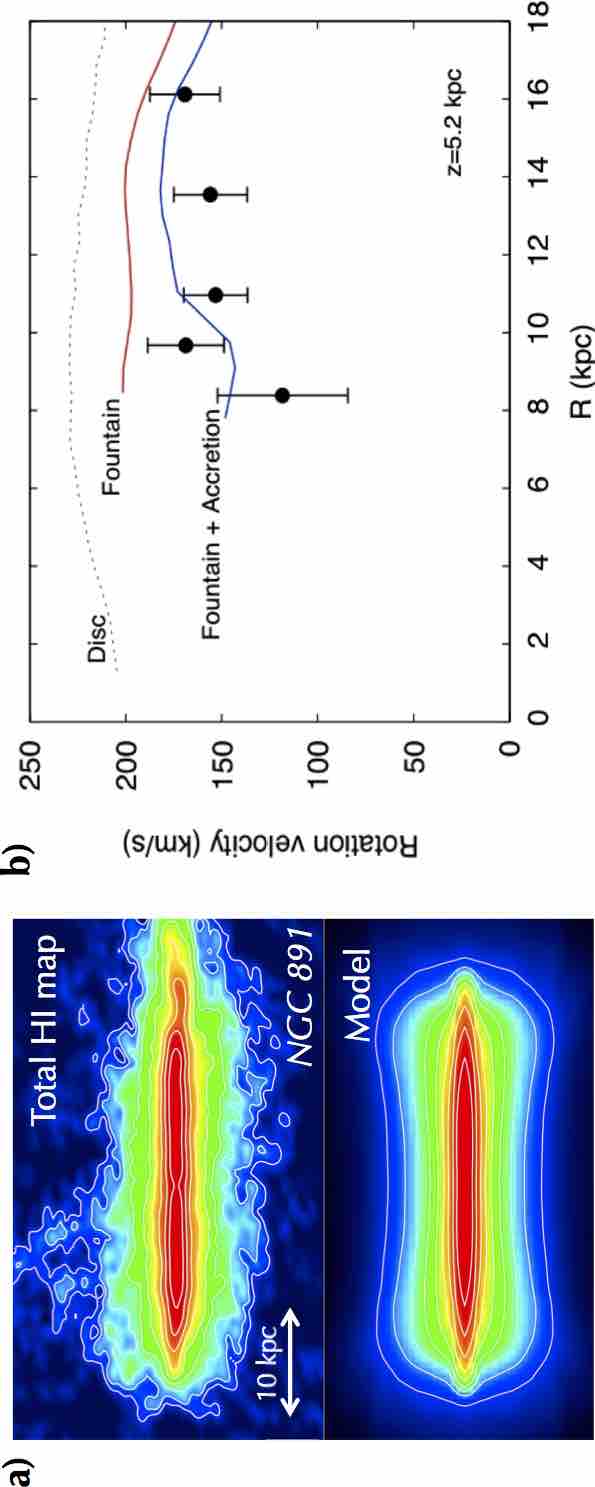}
\caption{
\emph{a)} Comparison between the total \hi\ map of NGC\,891 (rotated with its major axis along the x-axis) and the mock map produced by a galactic fountain model \citep[figure adapted from][]{Fraternali&Binney2006}.
The distribution of the gas in the extraplanar layer is reproduced by ejecting clouds from the disc with an average kick velocity of $v_{\rm kick}=80 \kms$.
\emph{b)} Comparison between the rotation above the plane of the extraplanar gas in NGC\,891 (at $z\sim 5 \kpc$) with a purely ballistic fountain (red curve) model and a fountain accretion (blue curve) model \citep{Fraternali&Binney2008}, see text.
}
\label{fig:fountainN891}
\end{figure}

Finally, the contribution of all the clouds in the simulation described above is recorded in a mock observation ($\hi$ \emph{pseudo-datacube}) of the model galaxy.
The comparison with real data is then carried out by building several models and by varying the parameters to minimize the data-model residuals. 
Figure \ref{fig:fountainN891}a shows the total $\hi$ map produced by one such galactic fountain model for the edge-on galaxy NGC\,891 (see also Fig.\ \ref{fig:ngc891}).
The model was produced by ejecting particles with $v_{\rm kick}\simeq 80 \kms$.
The distribution and the extent of the model extraplanar gas layer appears quite similar to the data \citep[see][for details]{Fraternali&Binney2006}.
However, the kinematics of the extraplanar gas predicted by this type of fountain model is not compatible with the data.
This can be clearly seen from Fig.\ \ref{fig:fountainN891}b where the observed rotation curve of the extraplanar gas at $z\sim 5 \kpc$ (points) is compared to the prediction of a ballistic fountain model (red curve).
The actual comparison was done much more thoroughly by simulating the whole datacube \citep{Fraternali&Binney2006}, but Fig.\ \ref{fig:fountainN891} makes the point quite straightforwardly. 
The rotation of the extraplanar gas predicted by the fountain model is systematically higher than the observed rotation.

This comparison concludes this description of the ballistic galactic fountain model.
The main feature of this model is that the particles conserve angular momentum throughout their trajectory.
However, Fig.\ \ref{fig:fountainN891} shows that their rotation velocities should decrease, very likely as a consequence of hydrodynamical effects that are not included in the ballistic treatment.
Unfortunately, a full hydrodynamical treatment of the problem is impractical because of the impossibility to fully cover the parameter space but it is possible to incorporate hydrodynamical effects on the motion of individual fountain clouds (\S \ref{sec:fountainAccretion}). 
These effects allow us to reconcile the galactic fountain model with the data (blue curve in Fig.\ \ref{fig:fountainN891}) and constitute the key evidence for gas accretion within the fountain cycle.
In \S \ref{sec:simulations}, we describe the tailored high-resolution hydrodynamical simulations that allowed us to go beyond the ballistic model.

\section{Hydrodynamical simulations of disc-corona mixing}
\label{sec:simulations}

The interface between galaxy discs and the surrounding environment is a very complex multi-phase region that must be studied using hydrodynamical simulations at very high resolution.
The standard approach is to simulated the whole galaxy disc or some fraction of it, either in isolation \citep{Melioli+2009,Marasco+2015} or in a full cosmological box \citep{Marinacci+2014,Schaye+2015}.
These studies allow us to study the global effect of feedback in the redistribution of gas \citep{Brook+2012}, the removal of gas from the galaxy \citep{Dave+2011,Pontzen&Governato2012} and the flow to and from the circumgalactic medium \citep{Shen+2013, Hobbs+2013}.
However, they suffer from the choice of the feedback recipes and the prescriptions used for a number of sub-grid phenomena that one must take into account.
Different groups and codes use various feedback prescriptions, often to maximize its efficiency, which are not in agreement with each other and can lead to very different results \citep{Scannapieco+2012}.
Moreover, the importance of phenomena often neglected in simulations like cosmic rays \citep{Girichidis+2016} and thermal conduction \citep{Keller+2014} is currently largely unconstrained.

\begin{figure}[!hb]
\includegraphics[angle=-90, clip=true, width=4.5in]{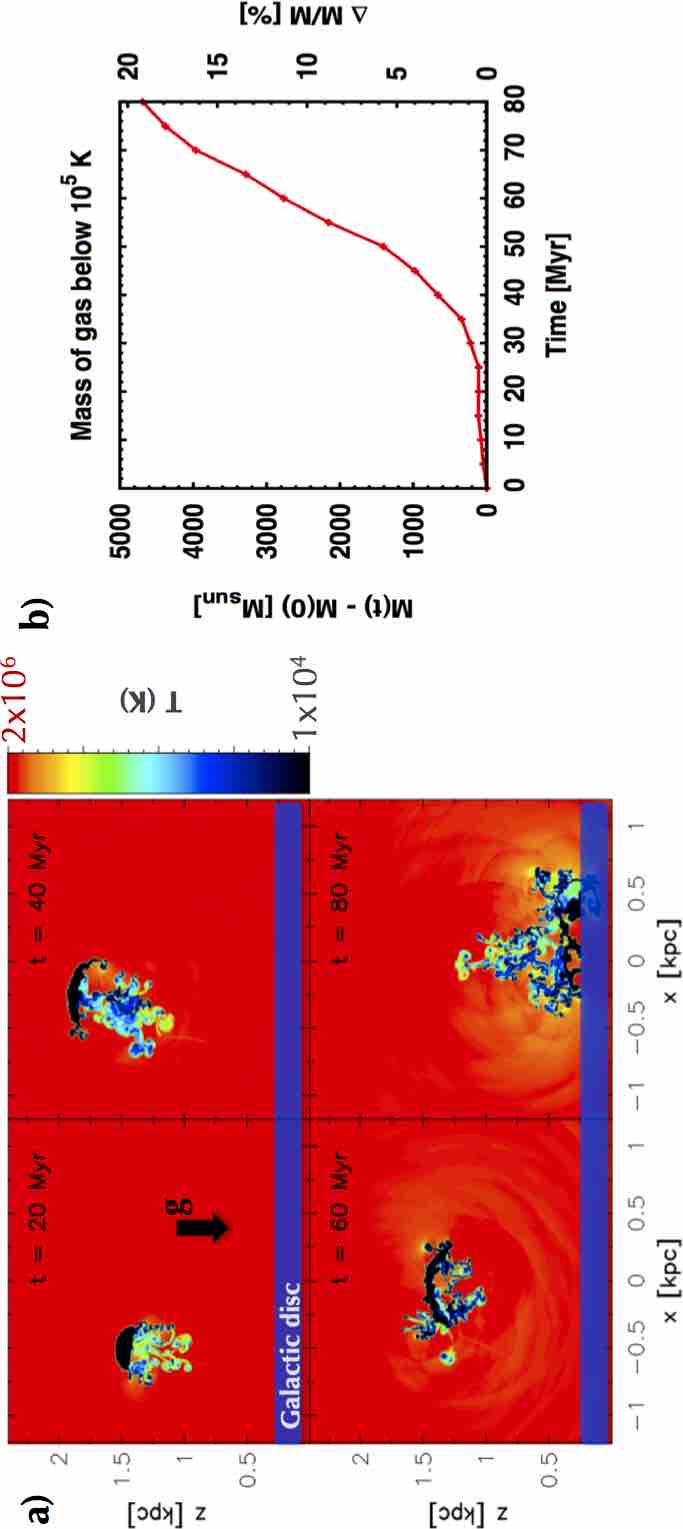}
\caption{\emph{a)} Temperature snapshots extracted at different times (see top right corners) from a 2D hydrodynamical simulation of a cold ($T=10^4 \K$) fountain cloud ejected upward from the Galactic plane and traveling through the hot ($T=2\times 10^6\K$) and metal-poor corona.
\emph{b)} Behavior of the mass of cold/warm ($T<10^5 \K$) gas in the simulation box. The mass increases with time because part of the corona condenses in the turbulent wake of the cloud.
This simulation is performed in 2D with the code ECHO++ \citep{Marinacci+2010} and with a grid size of $2\pc \times 2\pc$.
}
\label{fig:fountainSim}
\end{figure}

A complementary approach is to use targeted simulations at very high (parsec scale) resolution to study the gas mixing at the disc-corona interface \citep{Marinacci+2010, Marinacci+2011}.
With this approach, one does not aim to reproduce the entire phenomenon of supernova feedback but focuses on the physics of the interaction between disc and corona material.
In these simulations, the disc material ejected by the supernova feedback is represented by a \emph{cold} ($T\lesssim 10^4 \K$) cloud that travels through an ambient medium with properties typical of the galactic corona.
The cloud is typically at high metallicity (Solar or nearly so), while the corona is at lower metallicity usually 0.1 Solar, as observed in nearby galaxies \citep{Hodges-Kluck&Bregman2013, Bogdan+2013}.
Clouds and corona are initially in pressure equilibrium, but this assumption is not too stringent and experiments show that if one starts with different conditions, the equilibrium is promptly reached in times $\lesssim 10 \Myr$.
Figure \ref{fig:fountainSim} shows temperature snapshots of one such simulation \citep{Marinacci+2010}.
Here, a cold cloud with a mass of $\sim 10^4 \mo$ is ejected at a small angle from the normal to the galactic disc and travels through the corona.
The latter is a medium at $T=2\times 10^6\K$ in hydrostatic equilibrium with a vertically stratified galactic potential.

The fundamental result obtained with simulations like the one shown in Fig.\ \ref{fig:fountainSim}a is that the amount of cold gas in the box increases as a function of time (see Fig.\ \ref{fig:fountainSim}b). 
This happens because part of the coronal gas condenses in the turbulent wake of the cloud. 
This gas will then follow the cloud back to the disc providing fresh gas accretion, this is the essence of fountain-driven (or supernova-driven) accretion.

The condensation in the cloud's wake occurs for the following reasons.
The Milky Way's corona is at a temperature ($T\simeq 2\times 10^6\K$) and density ($n\simeq 10^{-3} \cmcu$, with $n$ total density, \citet{Anderson&Bregman2010}) at which the cooling time is relatively long:
\begin{equation}
t_{\rm cool} \simeq 
2.6 \times 10^9 \left(\frac{T}{2\times10^6\K}\right) \left(\frac{n}{10^{-3}\cmcu}\right)^{-1} 
\left[\frac{\Lambda(T,Z)}{\Lambda(2\times10^6\K,0.1\,Z_{\odot})} \right]^{-1}\yr
\end{equation}
where $\Lambda(T,Z)$ is the standard cooling function for collisional-ionization equilibrium \citep{Sutherland&Dopita1993} and we have used an electron density\footnote{Note that this is strictly valid only for $T>10^5\K$.} $n_{\rm e}\simeq 0.54n$.
The value of $\Lambda(T,Z)$ at $T=2\times10^6\K$ is low but it increases by almost two orders of magnitude from there to $T\simeq 2\times10^5\K$, roughly the peak of the cooling function.
In the wakes of the clouds, hot gas from the corona mixes with cooler gas from the disc thus $\Lambda(T,Z)$ moves progressively towards its peak.
Moreover, the disc gas is metal rich and this also increases the cooling rate of the mixture by up to another order of magnitude between 0.1 Solar and Solar.
In conclusion, the cooling time of the coronal material entrained in the turbulent wake decreases by orders of magnitudes making its condensation possible within the fountain cycle ($t_{\rm fount}\sim 100 \Myr$), see Fig.\ \ref{fig:fountainSim}b.

The main consequence of condensation is a steady transfer of material at the disc-corona interface from the hot to the cold phase, which then follows nearly ballistic orbits and rains down to the disc.
The average properties of the cold phase are significantly modified with respect to fountain orbits without condensation.
The first modification is the increase in mass: the added mass (Fig.\ \ref{fig:fountainSim}b) is made up exclusively of low-metallicity coronal material therefore providing significant accretion of metal-poor gas ($\sim 10-20\%$ of the initial mass of the fountain clouds). 
A second modification is introduced by the fact that the corona and the fountain gas have different angular momenta.
To visualize this, consider a corona in equilibrium that rotates with a velocity different from that of a typical fountain cloud:
\begin{equation}
v_{\rm \phi,\,hot}\ne v_{\rm \phi,\,cloud}\simeq v_{\rm c}(R,z)
\label{eq:vphiHot}
\end{equation}
where the equality is valid for a purely ballistic fountain (\S \ref{sec:fountain} and eq.\ \ref{eq:vc}).
Whenever $T_{\rm hot}$ is close to the virial temperature $T_{\rm vir}$, the velocity of the corona will be lower than the circular velocity ($v_{\rm \phi,\,hot}< v_{\rm \phi,\,cloud}$).
This can be intuitively understood by considering that such a corona is nearly sustained by pressure and it must give away part of its rotational support.
The condensation of material from a slowly rotating corona has then the crucial consequence of decreasing the global rotation velocity of the cold extraplanar gas and reconcile it with the observations (see Fig.\ \ref{fig:fountainN891}).
We see now how one can derive from the simulations the momentum transfer to be introduced in the fountain model.

\begin{figure}[ht]
\sidecaption
\includegraphics[trim=0cm 0cm 0cm 0cm, clip=true, angle=-90, width=\textwidth]{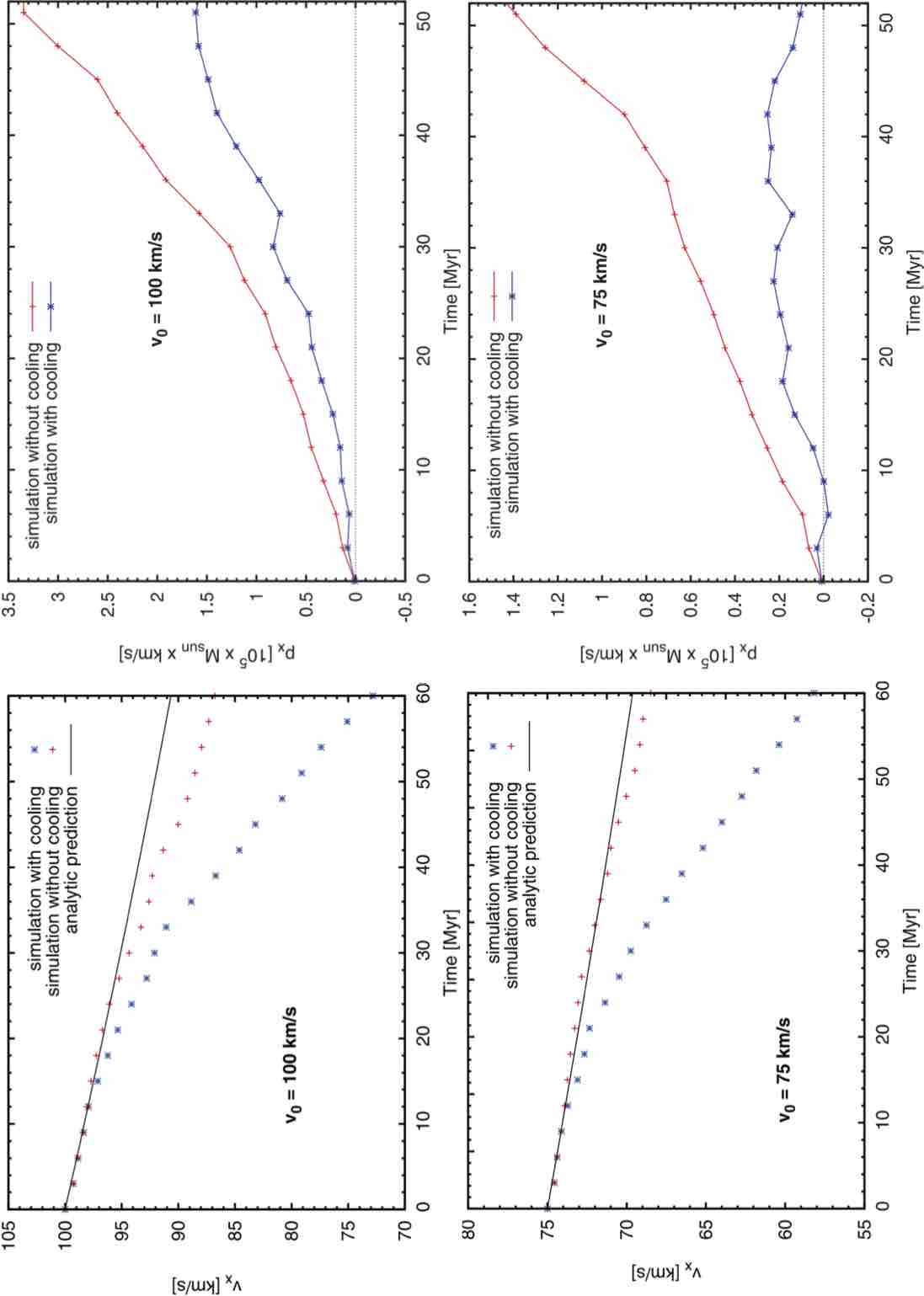}
\caption{Results of momentum transfer in simulations of fountain clouds that travel through a hot corona with initial relative velocities of $v=100\kms$ (top panels) and $v=75\kms$ (bottoms panels).
The left panels show the velocity of the cold ($T<3\times 10^4 \K$) gas as a function of time, while the right panels show the momentum acquired by the hot gas.
Red and blue curves refer to simulations in which the radiative cooling is, respectively, switched off (adiabatic) and on.
}
\label{fig:transfer}
\end{figure}

\begin{figure}[!ht]
\includegraphics[angle=-90, clip=true, width=4.5in]{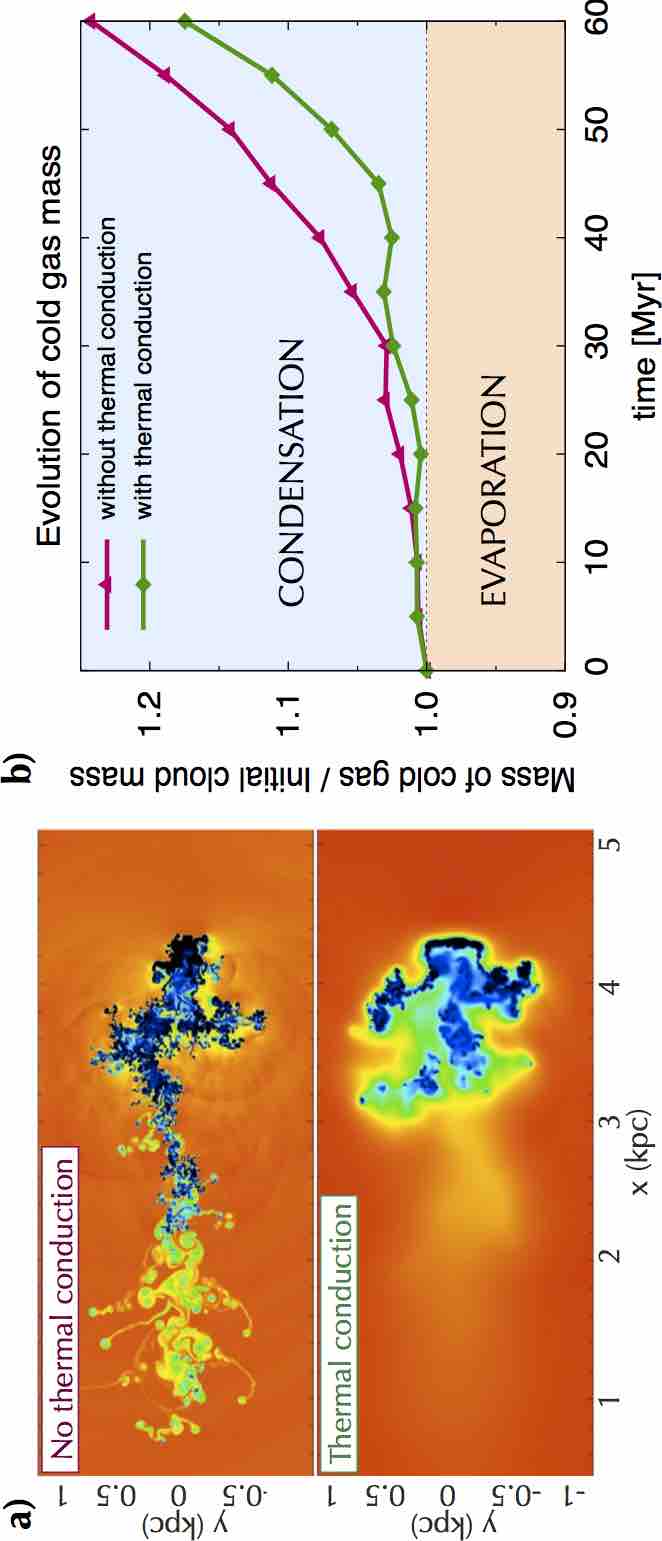}
\caption{
\emph{a)} Temperature snapshots extracted after $60 \Myr$ from a hydrodynamical simulation of a cold ($10^4 \K$) fountain cloud moving (along the x-axis from left to right) through a hot ($T=2\times 10^6\K$) and metal-poor ($Z=0.1\,Z_{\odot}$) medium, typical of the Milky Way's corona.
The top panel shows a simulation with radiative cooling akin to the one in Fig.\ \ref{fig:fountainSim}, while in the bottom panel thermal conduction is also included.
\emph{b)} Behavior of the mass of cold ($T\leq 2 \times 10^4 \K$) gas in the simulation box. The mass increases with time because of the condensation of the corona, even in the presence of thermal conduction.
These simulations were performed with the code Athena on a 2D grid with cell size $2\pc \times 2\pc$.
}
\label{fig:condensation}
\end{figure}

\citet{Marinacci+2011} used simulations like the one shown in Fig.\ \ref{fig:condensation} but without a gravitational field. 
They set up fountain clouds of masses $M_{\rm cold}={\rm few} \times 10^4 \mo$ moving through a hot medium with constant density. 
The neglect of gravity makes little difference in the property of the wake and the condensation while greatly facilitating the interpretation of the momentum transfer.
The stratification of the corona in $z$, induced by a gravity field, is very mild in the region where the fountain operates and the use of a constant density is also fully justified.
We note that fountain clouds have typical masses (see \S \ref{sec:fountainAccretion}) that make them fully pressure confined and self gravity plays no role \citep{Armillotta+2016}.
Fig.\ \ref{fig:transfer} (left panels) shows the velocity $v(t)$ of the cold ($T<3\times 10^4 \K$) gas in the simulation box as a function of time starting from initial relative velocities $v_0=100\kms$ and $v_0=75\kms$.
The velocity decreases because of two main effects: \emph{drag} and \emph{condensation}.
The drag can be easily parameterized as described in \S \ref{sec:fountainAccretion} and its effect is shown by the black curve.
This simple prescription, strictly valid for solids, works in fact quite well in predicting $v(t)$ in the adiabatic simulation.
However, when radiative cooling is turned on, part of the coronal material condenses in the cloud's wake and further reduces the velocity of the cold gas phase.
We discuss the parameterization of this condensation in \S \ref{sec:fountainAccretion}.

The right panels of Fig.\ \ref{fig:transfer} show the remarkable effect that condensation has on the hot ($T>1\times 10^6 \K$) phase.
Here we see the momentum acquired by the hot gas as a function of time.
Consider first the simulations without cooling, in this case the mass of the cold phase is slightly decreasing with time (not shown here) and so is its velocity, as a consequence the cold gas loses momentum.
This momentum is transferred to the hot gas, which had initially null momentum.
Only a negligible fraction of gas in these simulations occupies the regime at intermediate temperatures ($3\times 10^4 <T<10^6 \K$) and thus this does not contribute to the momentum budget.
This situation changes when the radiative cooling is included.
The velocity of the cold gas decreases much more rapidly but, as mentioned, its mass increases (Fig.\ \ref{fig:fountainSim}b). 
As a consequence, its momentum decreases less than in the case without cooling \citep{Marinacci+2011}.
This can be seen by looking at the momentum gained by the hot gas, which is significantly less than in the adiabatic case.
Moreover, now part of the momentum lost by the cold gas is transferred to material at intermediate temperatures, generated by the cooling of the corona.
All these effects conspire to have an approximately null transfer of momentum to the corona when the relative velocity between the cold and the hot gas falls below a threshold $v_{\rm thres}\simeq 75 \kms$ (Fig.\ \ref{fig:transfer}, bottom right).

The above findings have important consequences.
A problem that was identified in previous papers \citep{Fraternali&Binney2008} was that, given that the cold extraplanar gas has a mass larger than the lower corona (i.e.\ the hot gas within a few kpc from the disc, see Table \ref{tab:epg}), the transfer of momentum from the cold to the hot phase should spin the corona up to velocities so high that it would shoot away like a liquid in a centrifuge.
This is obviously not happening because we observe the corona in the central parts of the halo region (Fig.\ \ref{fig:ngc891}b).
The explanation is offered by the above result: there is a threshold in the velocity between extraplanar cold gas and the corona below which the hot gas does not gain momentum, therefore halting its spin-up.
In other words, the momentum transferred to the hot phase through ram pressure is given back to the cold phase by the cooling of the corona. 
Furthermore, this process furnishes a prediction of the rotational speed that we can expect the corona to have.
As long as the orbits of the fountain material are similar to the ones in Fig.\ \ref{fig:orbitsPure}, the largest velocity difference between corona and the fountain gas is expected to be in the azimuthal direction.
The vertical velocity contributes by about $\langle v_z\rangle\simeq 40 \kms$ on average, while the radial velocity can be safely neglected.
Thus, we can turn the momentum of the simulations in Fig.\ \ref{fig:transfer} into \emph{angular momentum} and its velocity into azimuthal relative velocity $v_{\rm lag}= \sqrt{v_{\rm thres}^2-\langle v_z\rangle^2}\simeq 63\kms$.
The threshold velocity changes slightly with the corona density, so for instance a density of $n=0.5\times 10^{-3} \cmcu$ ($n=2\times 10^{-3} \cmcu$) gives a $v_{\rm thres}\simeq 50 \kms$ ($v_{\rm thres}\simeq 85 \kms$ ) and $v_{\rm lag}\simeq 30\kms$ ($v_{\rm lag}\simeq 75\kms$).
These values are to be considered with respect to the extraplanar cold gas, which is already lagging by about $15 \kmskpc$ \citep{Oosterloo+2007, Marasco&Fraternali2011}.
In the end, we expect the lower corona (say at $z=1-2 \kpc$) to rotate with a lag of $v_{\rm lag}=45-105\kms$ with respect to the disc rotation\footnote{Note that the value of $80-120\kms$ quoted in \citep{Marinacci+2011} was obtained by considering only a density of $n=10^{-3} \cmcu$ and a corona extending up to $z=4\kpc$.}.
The rotation of the Milky Way's corona has been recently measured using O VII absorption lines and it was found to be $v_{\rm \phi,\,hot}=183\pm41 \kms$ if the disc rotates at $240 \kms$ and thus with a $v_{\rm lag}\simeq 60 \kms$ \citep{Hodges-Kluck+2016}, a value remarkably in line with the above prediction.

We conclude by remarking that, apart from predicting the rotation of the hot gas, the transfer of momentum from the cold to the hot phases and of mass from the hot to the cold phases has the major consequence of lowering the rotational velocity of the cold extraplanar gas.
This is exactly what it is needed to reconcile the fountain model with the observations (see \S \ref{sec:fountain} and Fig.\ \ref{fig:fountainN891}).
After a short digression on the role of thermal conduction, we see in \S \ref{sec:fountainAccretion} how this can be implemented in large-scale fountain models allowing us to estimate the gas accretion from the corona.

An important physical process that was missing from previous simulations \citep{Marinacci+2010, Marinacci+2011} is thermal conduction.
Thermal (or heat) conduction causes the direct transfer of heat from regions at different temperatures and it is particularly efficient where strong temperature gradients are present in small regions of space, something that occurs very often in the disc-corona interface.
In order to account for this process, \citet{Armillotta+2016} used the fixed-grid hydrodynamical code {\it Athena} \citep{Stone+2008} and modified it to include implicit and explicit treatments of thermal conduction.
They also include a standard radiative cooling function \citep{Sutherland&Dopita1993} and two metallicities for the different fluids.  
Here, we briefly discuss the effect of thermal conduction on coronal condensation.

Thermal conduction is implemented by introducing a heat flux that is described by the standard Spitzer formula \citep{Spitzer1962}:
\begin{equation}
\mathbf{F}_{\rm cond}=- f\, \kappa_{\rm Sp}\,T^{5/2}\,\grad T
\label{eq:spitzerTC}
\end{equation}
where T is the temperature at a given position, $\kappa_{\rm Sp}=1.84\times 10^{-5}\ln\Psi \ergs \K^{-1} \cm^{-1}$ and the Coulomb logarithm is $\ln\Psi\simeq 30$ with a small dependence on temperature.
The factor $f$ in front of eq.\ \ref{eq:spitzerTC} is set to $f=0.1$ and takes into account the suppression of conduction by magnetic fields \citep[see][for details]{Armillotta+2016}.
Whenever the temperature scale length ($T/\grad T$) becomes shorter than the free mean paths of the electrons, the Spitzer formula is not valid any longer thus a prescription for the so-called saturated conduction \citep{Dalton&Balbus1993} was also implemented.

The general effect of thermal conduction is to smooth the temperature gradients at the interface between cold (fountain) gas and the hot medium of the corona.
This has two main consequences.
First, it makes the cloud more compact and resilient to destruction via Kelvin-Helmholtz instabilities \citep{Vieser&Hensler2007}.
Second, small cloudlets extracted from the main body of the cloud evaporate quite easily.
As a result, the cloud's wake is mostly made of material at intermediate temperatures rather than very cold knots.
This can be seen in Fig.\ \ref{fig:condensation}a where we show temperature snapshots of simulations of clouds moving through the corona along the horizontal axis.
These snapshots are taken after $60 \Myr$ and apart from the switching on/off of thermal conduction they share the same initial setup.
The right panel of Fig.\ \ref{fig:condensation} shows the mass of cold ($T\leq 2\times 10^4\K$) gas in the simulation box as a function of time.
The rate of condensation is lower in the simulation with thermal conduction.
However, in the time of a typical fountain cycle (see \S \ref{sec:fountain}), the mass of cold gas still increases substantially and thus we should expect significant gas accretion from the corona to occur in the region where the fountain operates.

\section{Galactic fountain with accretion: beyond the ballistic model}
\label{sec:fountainAccretion}

The simulations presented in \S \ref{sec:simulations} show that the interaction between the galactic fountain and the hot gas can be described as a composition of drag (ram pressure) and condensation.
The deceleration of the fountain cloud due to the drag can be easily parameterized as:
\begin{equation}
\mathbf{a}_{\rm drag}=-\frac{v^2}{v_0 t_{\rm drag}}\frac{\mathbf{v}}{v}
\end{equation}
where $\mathbf{v}$ is the relative velocity between the cloud and the hot gas, $v_0$ is the initial relative velocity. The \emph{drag time} is:
\begin{equation}
t_{\rm drag}=\frac{M_{\rm cloud}}{C\rho_{\rm hot}\sigma_{\rm cloud} v_0}
\end{equation}
where $M_{\rm cloud}$ and $\sigma_{\rm cloud}$ are the mass and the cross section of the cloud, $\rho_{\rm hot}$ is the density of the corona and $C$ is a dimensionless constant of order unity to account for the geometry of the cloud. 
Assuming that mass and cross section do not change during the cloud motion, the velocity of the cold gas decreases as follows:
\begin{equation}
v(t)=\frac{v_0}{1+t/t_{\rm drag}}.
\label{eq:vt}
\end{equation}
Equation \ref{eq:vt} gives the black curves shown in the left panels of Fig.\ \ref{fig:transfer} and it is a good prediction for simulations without radiative cooling.

\begin{figure}[ht]
\sidecaption
\includegraphics[trim=0cm 0cm 0cm 0cm, clip=true, angle=0, width=0.52\textwidth]{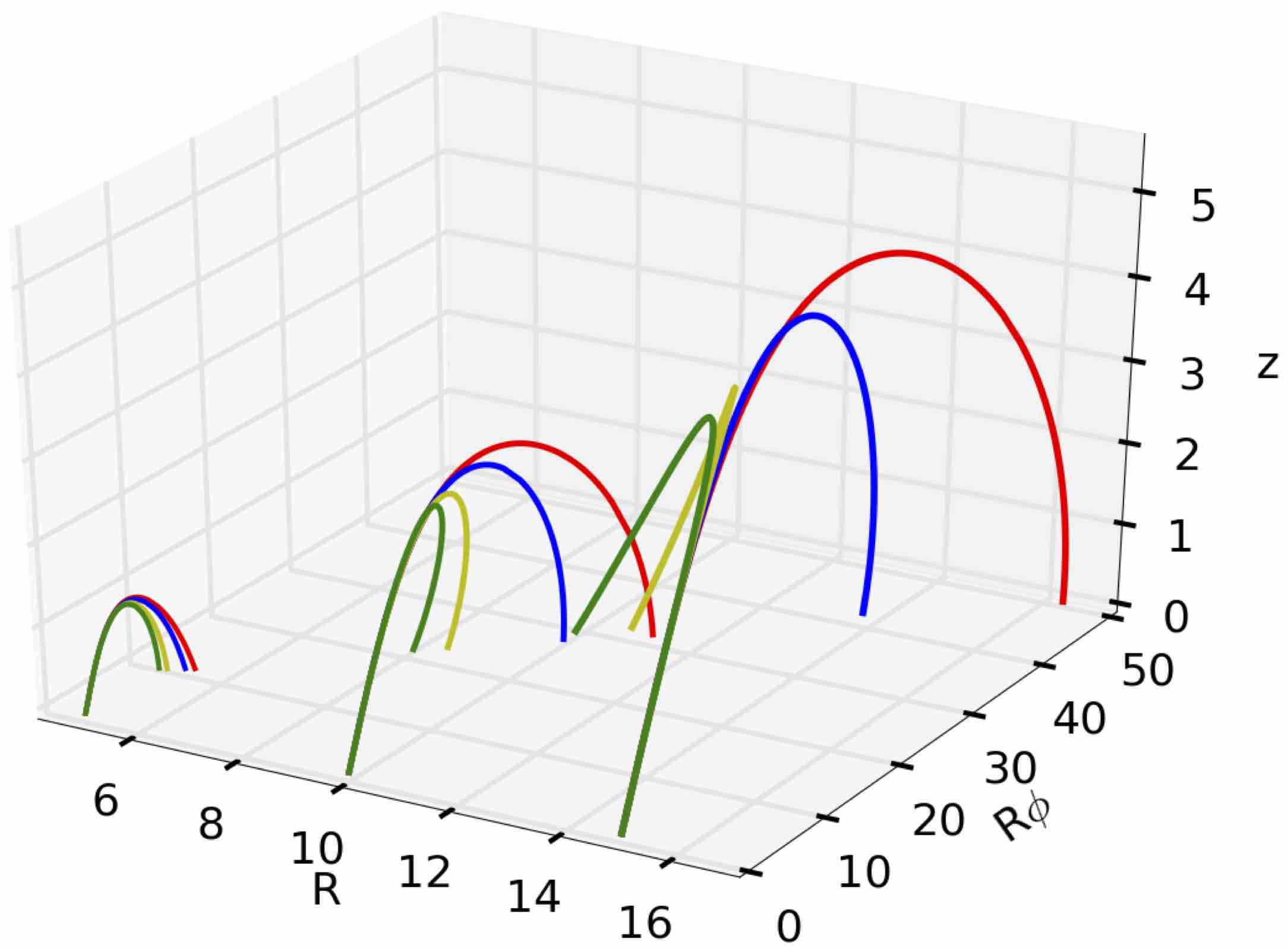}
\includegraphics[trim=0cm 0cm 0cm 0cm, clip=true, angle=0, width=0.47\textwidth]{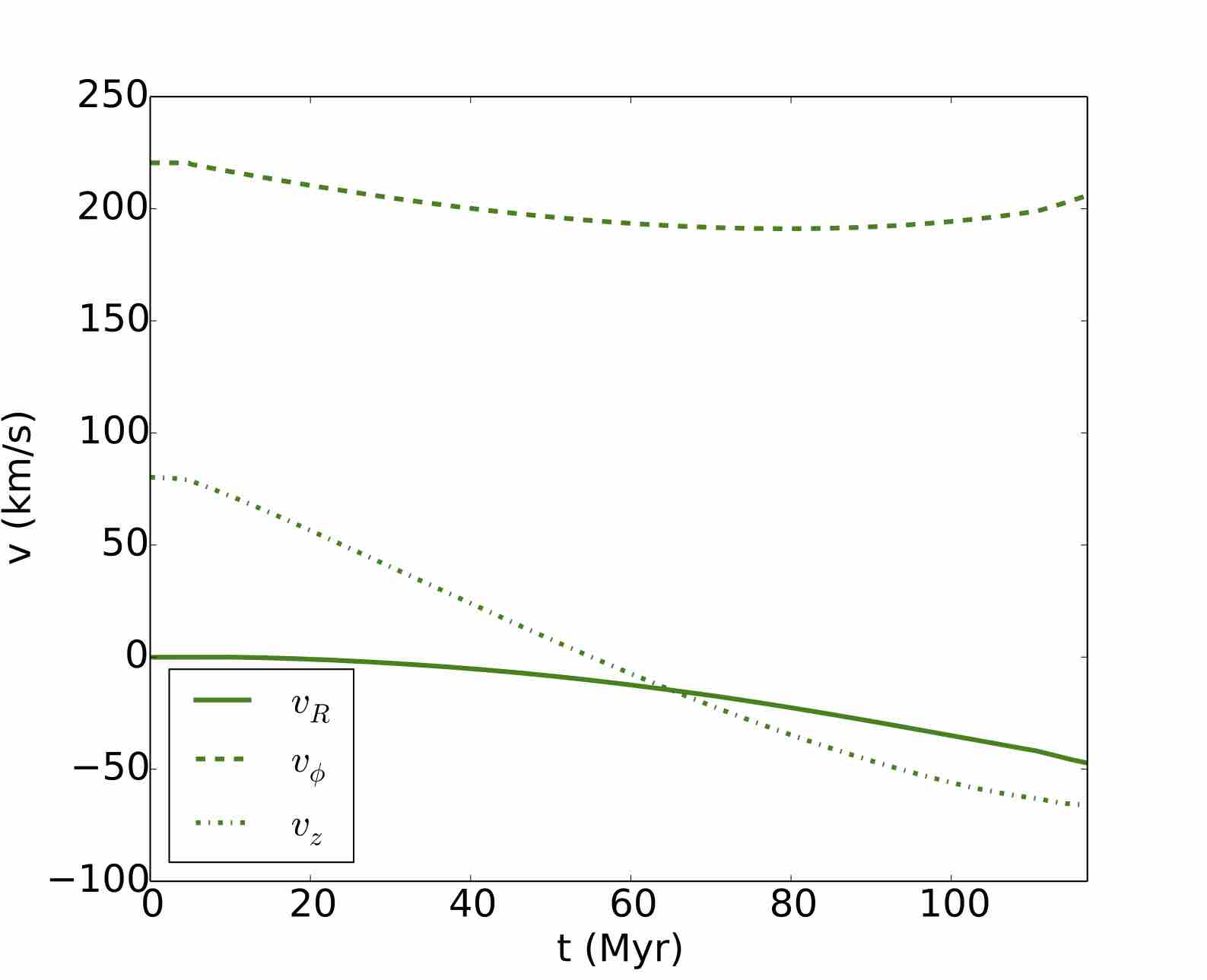}
\caption{
\emph{a)} Orbits of galactic fountain particles in the potential of the Milky Way shot vertically upwards with $v_{\rm kick}=75 \kms$ from three radii $R=5,\,10,\,15 \kpc$.
The various effects of drag alone (blue), condensation/accretion alone (pea green) and drag+accretion (dark green) are shown and compared with the pure fountain (red, see also Fig.\ \ref{fig:orbitsPure}).
\emph{b)} Cylindrical components of the velocity of the fountain particle ejected from $R=10 \kpc$ when drag and accretion are taken into account.
}
\label{fig:orbitsAccretion}
\end{figure}

The accretion of coronal gas into the fountain wake has been described both as an exponential or a power-law growth \citep{Fraternali&Binney2008, Fraternali+2015}.
Here we use an exponential growth for simplicity, thus the mass ($M$) of the cold phase grows as:
\begin{equation}
\dot M=\alpha M
\label{eq:mdot}
\end{equation}
where $\alpha$ is a \emph{condensation parameter} to be determined, inverse of the condensation timescale.
At each infinitesimal time step ($\delta t$) a mass $\delta M$ from the corona joins the cold phase.
The phenomenon can be described as an inelastic collision and the conservation of momentum gives a velocity for the cold gas at $t+\delta t$ \citep{Fraternali&Binney2008}:
\begin{equation}
v(t+\delta t)=\frac{M(t) v(t)+\delta M v_{\rm hot}}{M(t)+\delta M}
\end{equation}
where $v_{\rm hot}=0$ if we are considering relative velocities.
Then, given that from eq.\ \ref{eq:mdot} $\delta M=\alpha M \delta t$, the deceleration due to condensation simply becomes:
\begin{equation}
\mathbf{a}_{\rm accr}=-\alpha \mathbf{v}.
\end{equation}
and thus $v$ decreases exponentially at the same rate at which the mass increases.
Figure \ref{fig:orbitsAccretion} shows the effect of the inclusion of drag and condensation/accretion in the orbits of fountain clouds in the Milky Way's potential.
The clouds are shot vertically from three radii ($R=5,\,10,\,15 \kpc$).
The main effect of drag and accretion is the reduction of the specific angular momentum (and vertical momentum) of the cold gas.
This has the following consequences: 1) the cold gas reaches lower heights (if shot with the same $v_{\rm kick}$), 2) its $v_{\phi}$ decreases more rapidly than in a pure fountain and 3) it tends to move inward thus acquiring a negative radial velocity.
As we see in \S \ref{sec:observations}, these effects are easily detectable in the kinematics of the extraplanar $\hi$ and therefore the values of the parameters of the models can be accurately determined.

In conclusion, the parameters of a galactic fountain model that includes hydrodynamical effects of the interaction between the fountain gas and the corona can be reduced to the following:
i) the characteristic kick velocity $v_{\rm kick}$, 
ii) the ionized fraction in the ascending part of the orbit $f_{\rm ion}$ (\S \ref{sec:fountain}),
iii) the drag time $t_{\rm drag}$,
iv) the condensation/accretion rate $\alpha$.
In principle, all these parameters can have spatial variations although in most models until now they have been kept constant for simplicity.
The drag time can be reasonably fixed by considering a typical mass for the fountain clouds of $M_{\rm cloud}> 1 \times 10^4 \mo$ \citep{Marinacci+2010}.
Indeed Galactic Intermediate-Velocity Clouds (IVCs) and HVCs have typical massed between a few $\times 10^4 \mo$ and a few $\times 10^6 \mo$, whereas smaller masses are unlikely to survive for long time in the corona \citep{Heitsch&Putman2009, Kwak+2011}.
If the density does not change significantly within the cloud, its size can be simply estimated as:
\begin{equation}
R_{\rm cloud}= \left(\frac{3}{4 \pi} \frac{T_{\rm cloud}\, \mu_{\rm hot}\, M_{\rm cloud}}{T_{\rm hot}\, \mu_{\rm cloud}\, \rho_{\rm hot}} \right)^{1/3}
\end{equation}
where $\mu_{\rm cloud}$ and $\mu_{\rm hot}$ are the atomic weights and we have imposed pressure equilibrium between cloud and corona ($n_{\rm hot}T_{\rm hot}=n_{\rm cloud}T_{\rm cloud}$).
The drag time becomes:
\begin{equation}
t_{\rm drag}\simeq 570 \left(\frac{M_{\rm cloud}}{10^4 \mo}\right)^{1/3}
\left(\frac{n_{\rm hot}}{10^{-3} \cmcu}\right)^{-1/3}
\left(\frac{T_{\rm hot}/T_{\rm cloud}}{200}\right)^{2/3}
\left(\frac{v_0}{75 \kms}\right)^{-1} \Myr.
\label{eq:tdrag2}
\end{equation}
for a neutral cloud, a fully ionized hot medium and assuming $C=1$.
As mentioned, we can expect that for most clouds $M_{\rm cloud}>10^4 \mo$ and $t_{\rm drag}$ in eq.\ \ref{eq:tdrag2} to be a lower limit.
Thus in general, the drag time is larger than the orbital time ($t_{\rm fount}\sim 100 \Myr$) and we can conclude that, as shown by simulations (Fig.\ \ref{fig:transfer}), the main effect on the orbit is due to the condensation of coronal gas.

\section{Observational evidence of fountain-driven accretion}
\label{sec:observations}

The observational benchmark for the fountain accretion model described in this Chapter has been the Milky Way, where several tracers have been used to test its validity and estimate the gas accretion from corona condensation.
In this Section we describe these successful tests in some detail.
We first briefly mention early studies carried out by Fraternali \& Binney (2006, 2008) on the two external galaxies NGC\,891 and NGC\,2403 \citep{Fraternali&Binney2006, Fraternali&Binney2008}.
The main feature of the extraplanar gas of NGC\,891 is the lag in rotation that is not reproduced by a simple galactic fountain model (Fig.\ \ref{fig:fountainN891}b).
The introduction of accretion onto fountain clouds following eq.\ \ref{eq:mdot} produces the nice match with the data shown by the blue curve in Fig.\ \ref{fig:fountainN891}.
This fit to the data requires an accretion of $\dot M_{\rm acc}\simeq 3 \moyr$, slightly lower than the galaxy's SFR of $3.8 \moyr$ \citep{Popescu+2004}.

The second galaxy that was considered in these early works is NGC\,2403, a smaller galaxy with relatively high SFR=$1.2 \moyr$ \citep{Kennicutt+2003} and large amounts of extraplanar gas \citep{Fraternali+2002}.
The kinematics of the extraplanar $\hi$ is characterized by a lagging rotation and a global inflow towards the center. 
Remarkably, the fountain accretion models is able to reproduce this pattern very well\ \citep{Fraternali&Binney2008}.
The inflow is essentially due to the acquisition of low angular momentum from the ambient gas (Fig.\ \ref{fig:orbitsAccretion}).
In order to reproduce the observed kinematics of the extraplanar gas, accretion at a rate of about $\dot M_{\rm acc}\simeq 0.8 \moyr$ is required.
We stress that these accretion rates are estimated through the kinematics of the extraplanar gas that has no direct relation with the SFR except through the fountain model.
Thus, the fact that the accretion rates are found be similar the SFRs is a strong confirmation of the soundness of the fountain accretion model (or an extraordinary coincidence).

In more recent years, the fountain accretion model was used to reproduce the $\hi$ emission of the extraplanar neutral gas in the Milky Way \citep{Marasco&Fraternali2011}.
This component had been studied in several papers before mostly in the form of individual clouds, IVCs or HVCs \citep{Richter+2001,vanWoerden+2004}.
The general consensus is that IVCs, being at relatively small anomalous velocities and at nearly Solar metallicity, are fountain clouds, whereas HVCs are more likely external clouds. 
This separation is now rather debated and it may be less neat than previously thought.
IVCs are just a local manifestation of a much broader component that extends above and below the plane and constitutes the extraplanar layer of the Milky Way (see table \ref{tab:epg}) \citep{Marasco&Fraternali2011}.
HVCs, to some extent, may be produced by extreme and rare events of the same kind that produced IVCs (see below).
Clearly, the first question to address in this context is whether the extraplanar $\hi$ is reproduced by a fountain model and whether its kinematics requires interactions with the galactic corona.

\begin{figure}[ht]
\begin{center}
\sidecaption
\includegraphics[trim=0cm 0cm 0cm 0cm, clip=true, angle=-90, width=\textwidth]{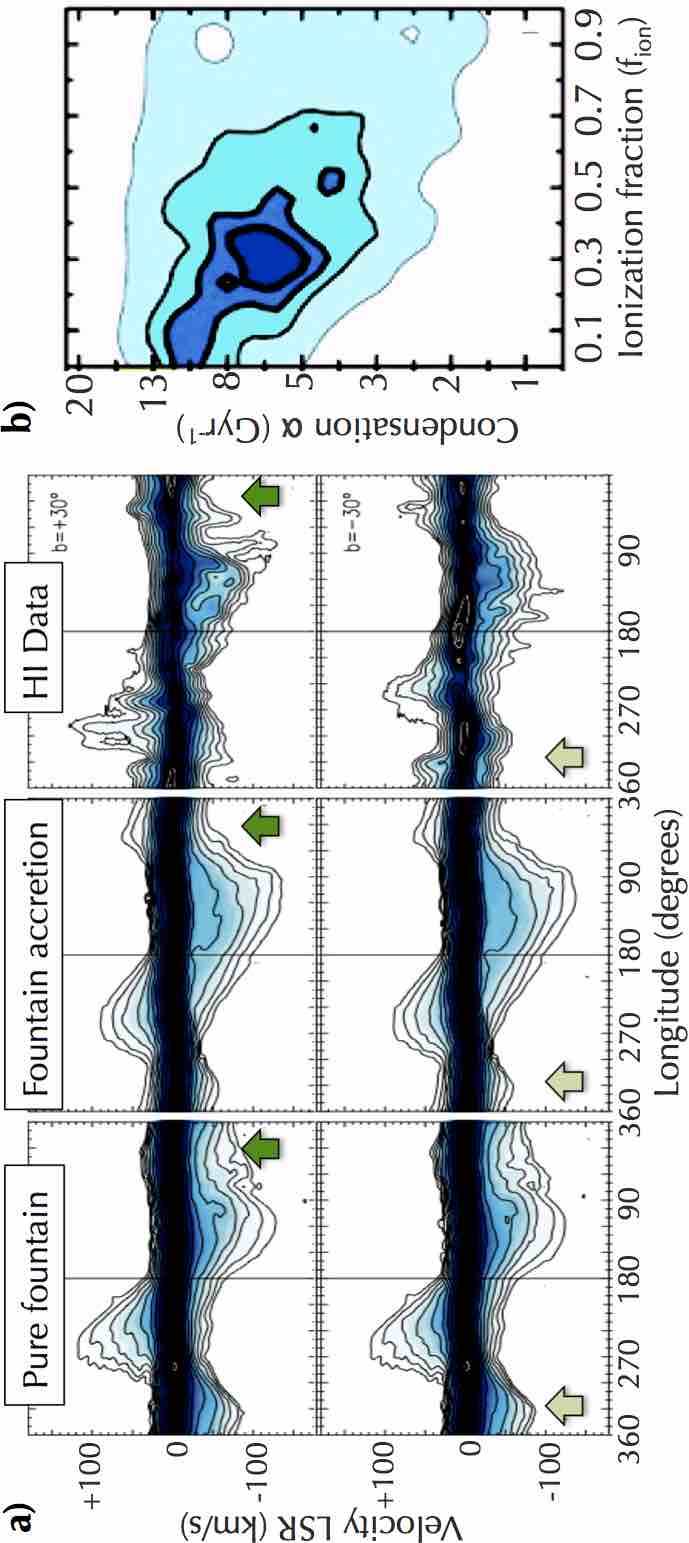}
\caption{
\emph{a)} Comparison between the \hi\ emission predicted by two fountain models and the LAB data (third column) in longitude-velocity plots extracted at two latitudes (top: $b=30^{\circ}$; bottom: $b=-30^{\circ}$).
The first column shows the best-fit pure galactic fountain and the second the best-fit fountain accretion model. 
The arrows indicate regions where the accretion model reproduces the data better than the pure fountain.
\emph{b)} Confidence levels of residuals between models with corona condensation and data evaluated in the parameter space ($\alpha$,$f_{\rm ion}$) for $v_{\rm kick}=70\kms$.
The minimum corresponds to a global $\hi$ accretion rate onto the Milky Way's disc $\dot M_{\rm acc}=1.6 \moyr$.
}
\label{fig:labFit}
\end{center}
\end{figure}

To answer this question, \citet{Marasco+2012} built a model of galactic fountain that included the effect of drag and condensation (see \S \ref{sec:fountainAccretion}).
This model predicts the $\hi$ emission from fountain clouds that can be compared with the datacube from the Leiden-Argentina-Bonn (LAB) survey \citep{Kalberla+2005}.
The comparison is carried out by building a suite of all-sky pseudo-datacubes with different values of the main parameters and calculating the residuals between models and data in the 3D-space (position on the sky and line-of-sight velocity).
Only locations where the extraplanar emission is present are used for the minimization (i.e.\ the Galactic disc is excluded).
Figure \ref{fig:labFit}a shows the comparison between models with and without coronal condensation and the $\hi$ data.
Data and models are shown as longitude-velocity plots extracted at two representative latitudes above and below the plane ($b=\pm30 \deg$).
At several locations, one can appreciate that the inclusion of gas accretion from the corona improves the fit substantially (see arrows).
These are, in particular, locations where radial inflow motions, typical of an interacting fountain (Fig.\ \ref{fig:orbitsAccretion}), are present.

Figure \ref{fig:labFit}b shows the values of the residuals between data and models obtained varying the two main parameters of the galactic fountain: the condensation parameter $\alpha$ and the ionization fraction $f_{\rm ion}$.
A clear minimum is found for $f_{\rm ion}=0.3$ (the gas is ionized for the initial $30\%$ of the ascending part of the orbits) and $\alpha=6.3 \Gyr^{-1}$.
This latter corresponds to a condensation/accretion timescale of $t_{\rm acc}=160\Myr$, which then translates into a global accretion rate of $\dot M_{\rm acc}\simeq 2 \moyr$ including helium.
In performing this fit the condensation timescale was not fixed to the value given by hydrodynamical simulations (Fig.\ \ref{fig:transfer}).
However, it remarkably turned out to be compatible with what the simulations predict \citep[details in][]{Marasco+2012}. 
Note also that, as expected, it is shorter than the drag time (eq.\ \ref{eq:tdrag2}).
The model predicts that nearly $20-25\%$ of the extraplanar $\hi$ in the Milky Way is made up of gas condensed from the corona, while the other $75-80\%$ is fountain material.
Individual clouds (IVCs/HVCs) may have different ratios depending on their past trajectories.

\begin{figure}[ht]
\begin{center}
\includegraphics[angle=-90, clip=true, width=4.5in]{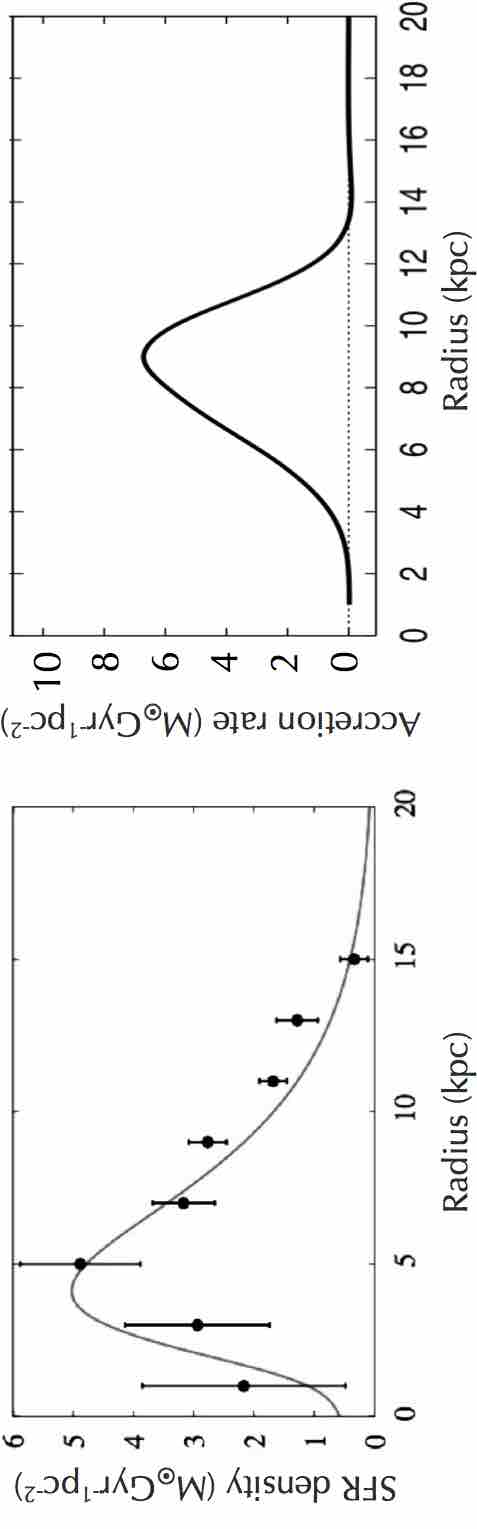}
\caption{
\emph{Left:} Star formation rate surface density in the Milky Way from supernova remnants \citep{Case&Bhattacharya1998}; figure from \citet{Pezzulli&Fraternali2016}.
\emph{Right:} Accretion rate surface density from corona condensation predicted by the fountain-driven accretion model; figure from \citet{Marasco+2012}.
}
\label{fig:rates}
\end{center}
\end{figure}

The application of the fountain accretion model to the Milky Way is successful at reproducing the detailed kinematics of the extraplanar $\hi$ and it returns a global accretion rate of the order of the Galactic SFR (Table \ref{tab:evidence}). 
We now discuss an interesting prediction of this model.
Figure \ref{fig:rates} shows a comparison between the SFR surface density in the Milky Way (left) and the gas accretion surface density (right) predicted by the fountain accretion model.
Clearly, most of the accretion tends to occur at larger radii with respect to the peak of the SFR density.
This is a consequence of the way the fountain operates: on the one hand, the orbital paths increase with radius (see Fig.\ \ref{fig:orbitsAccretion}) and the gas mass accreted from the corona increases with the orbital path (eq.\ \ref{eq:mdot}).
On the other hand, the fountain flow (ejection rate density) has a shape proportional to the SFR density.
The combined effect of the above is the accretion rate in Fig.\ \ref{fig:rates} (right panel).
The fate of the gas accreted from the corona at large radii is to move inward (before forming stars) because it has a rotation velocity lower than the local circular velocity (eq.\ \ref{eq:vphiHot}).
We discuss this further in \S \ref{sec:evolution}.

\begin{figure}[ht]
\begin{center}
\sidecaption
\includegraphics[trim=0cm 0cm 0cm 0cm, clip=true, angle=-90, width=\textwidth]{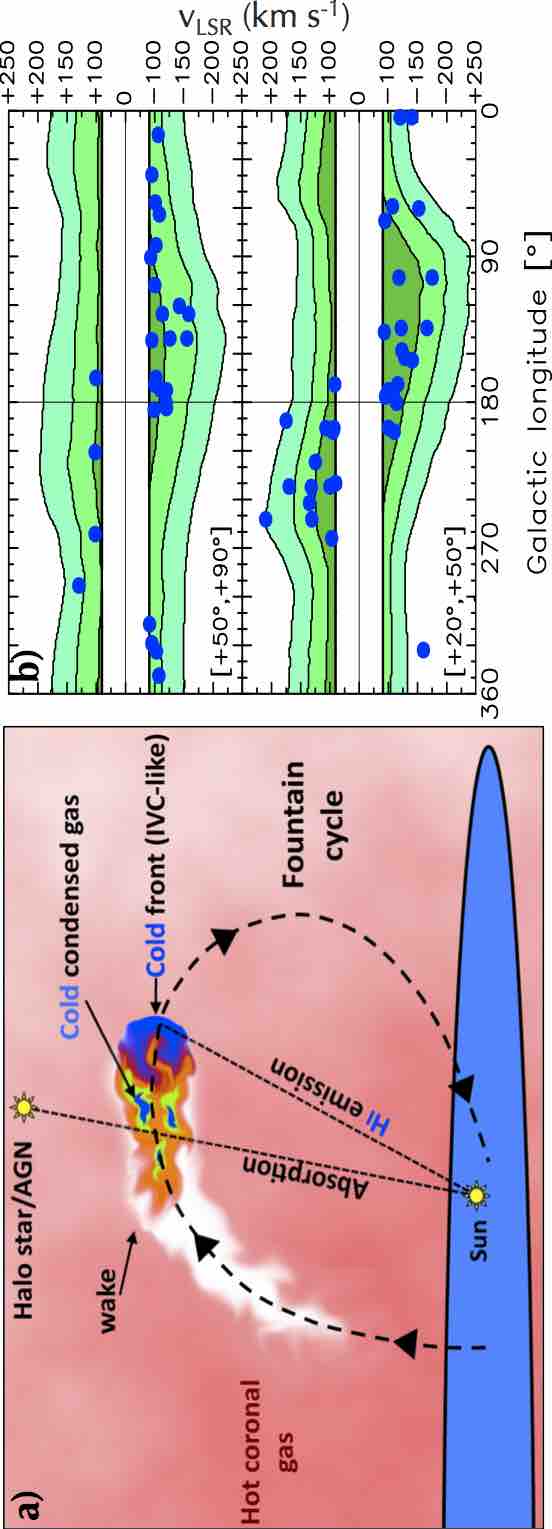}
\caption{
\emph{a)} 
A sketch of a fountain cloud, ejected from the disc by supernova feedback and interacting with the hot gas in the corona. 
In the turbulent wake, coronal gas mixes with high-metallicity disc material triggering the condensation of a fraction of the corona, which is then accreted onto the disc. 
An observer looking toward a background source intercepting the wake detects absorption lines from the ionized intermediate-temperature material. 
An observer looking toward the cold front detects $\hi$ emission at velocities typical of IVCs and, occasionally, HVCs.
\emph{b)} 
Predicted locations (darker shades show higher probability regions) in longitude and velocity of the intermediate-temperature material based on the fountain accretion model.
The points show the HST/COS absorption features detected in the temperature range $4.3<\log(T/\K)<5.3$.
The vast majority ($94\%$) of this ionized gas is consistent with being generated in wakes of fountain clouds.
}
\label{fig:sketchAntonino}
\end{center}
\end{figure}

The cooling of a significant fraction of the Galactic corona in the region close to the disc must also produce a large amount of gas at temperatures intermediate between the virial temperature and that of the cold $\hi$ gas.
Material at these temperatures is indeed observed surrounding the Milky Way's disc thanks to the detection of absorption features towards quasars \citep{Sembach+2003,Shull+2009}.
Recently, these absorbers have been identified also in the spectra of halo stars, which have the advantage to give an upper limit to the distance of the gas 
\citep[the estimated distance of the star][]{Lehner&Howk2011}.
\citet{Fraternali+2013} compared the predictions of the fountain accretion model with observations from the HST Cosmic Origins Spectrograph (COS) of low-ionization absorbers like \siiii, \siiv, \cii, \ciii\ and \civ, which encompass the temperature range $4.3<\log(T/\K)<5.3$ \citep[see also][]{Marasco+2013}.
Figure \ref{fig:sketchAntonino}a shows a sketch of the physical process.
The intermediate temperature material arises in the turbulent wakes of fountain clouds, where coronal condensation occurs.
The properties of the intermediate-temperature material with respect to the $\hi$ (velocity lag and relative density) were derived from hydrodynamical simulations like those in \S \ref{sec:simulations}.
Figure \ref{fig:sketchAntonino}b shows the areas in the longitude-velocity space where the fountain model predicts the presence of intermediate-temperature gas compared to the data \citep[points,][]{Lehner+2012}.
A statistical test shows that $94\%$ of the detected low-ionization absorbers are consistent with being produced in fountain wakes. 
The model also predicts the observed number of features along a typical line of sight and explains the large line broadening as due to turbulence in the wake.
We stress that to obtain these results, the parameters of the fountain accretion model were not fit to the COS data but kept fixed to those (see Table \ref{tab:evidence}) that reproduce the kinematics of the $\hi$ (Fig.\ \ref{fig:labFit}).

\begin{table}
\caption{Observational evidence of fountain-driven accretion}
\begin{center}
\begin{tabular}{lccccccc}
\hline\noalign{\smallskip}
Object & Tracer & DATA & SFR & $v_{\rm kick}$ & $f_{\rm ion}$ & Accretion rate & Ref. \\
 & & & ($\moyr$) & ($\kms$) & & ($\moyr$) & \\
\noalign{\smallskip}\svhline\noalign{\smallskip}
NGC\,891$^a$ & extraplanar $\hi$ & WSRT & 3.8 & 80 & $0-1$ & $\sim 3$ & (a) \\ 
NGC\,2403$^a$ & extraplanar $\hi$ & VLA & 1.2 & 55 & 0 & 0.8 & (a) \\    
Milky Way & extraplanar $\hi$ & LAB & $1-3$ & 70 & 0.3 & $\sim 2$ & (b) \\    
Milky Way & Metal absorbers & HST/COS & $1-3$ & 70$^b$ & 0.3$^b$ & $\sim 1$ & (c)  \\
HVC complex C & $\hi$+metals & LAB+COS& -- & 210 & -- & $\sim 0.25^c$ & (d) \\
\noalign{\smallskip}\hline\noalign{\smallskip}
\end{tabular}\\
\end{center}
$^a$ In these early works the ambient accreting medium was modeled as infalling and non-rotating.
$^b$ These parameters were not fit to the COS data but kept fixed to the values obtained from the $\hi$ fit.
$^c$ This is the total accretion rate if complex C keeps condensing at an exponential rate and all its warm/cold gas makes it back to the disc. References: (a) \cite{Fraternali&Binney2008}; (b) \cite{Marasco+2012}; (c) \cite{Fraternali+2013}; (d) \cite{Fraternali+2015}.
\label{tab:evidence}
\end{table}

\begin{figure}[ht]
\begin{center}
\sidecaption
\includegraphics[trim=0cm 0cm 0cm 0cm, clip=true, angle=-90, width=\textwidth]{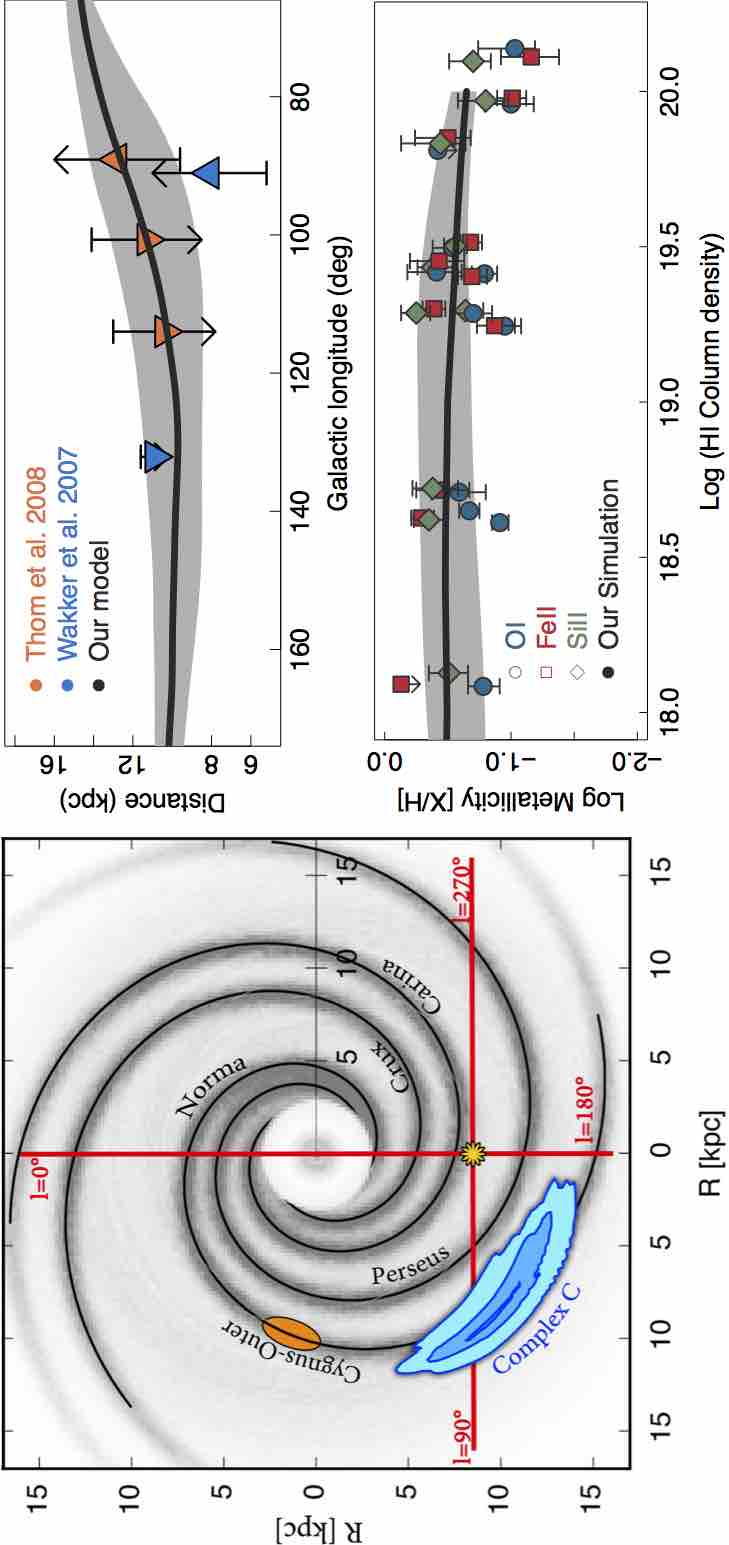}
\caption{
\emph{Left:} 
Location of the HVC complex C seen from a face-on view. 
Contours are obtained from a fountain accretion model and correspond to $\hi$ column densities of 1, 2 and 3 $\times 10^{19} \cmsq$ \citep{Fraternali+2015}. 
The spiral arms, the position of the Sun and the location of the superbubble blowout (orange ellipse) of the seed material are shown.
\emph{Right:}
Comparison between the distance from the Sun as a function of longitude (top) and the abundances of oxygen, iron and silicon as a function of $\hi$ column density (bottom) in the data (points and limits) and in hydrodynamical simulations of the complex C (black curve). The gray bands show the 16 and 84 percentiles. Adapted from \citet{Fraternali+2015}.
}
\label{fig:complexC}
\end{center}
\end{figure}

We now turn to the Galactic HVCs \citep{Wakker&vanWoerden1997}, whose origin has been debated since their discovery with the two competing scenarios being either gas accreting from the intergalactic medium into the Milky Way \citep{Oort1970} or galactic fountain material \citep{Bregman1980}.
The estimate of a low metallicity for the prototypical cloud complex C \citep{Wakker+1999} pointed strongly at an external origin but the source of the accretion remained unknown.
With the exclusion of spontaneous thermal instabilities in the corona \citep{Binney+2009}, others possibilities are gas from satellites \citep{Olano2008} or accretion from cosmological filaments \citep{Fernandez+2012}.
\citet{Fraternali+2015} investigated whether complex C could have originated by the blowout of a powerful superbubble in the disc of the Milky Way that seeded the condensation of a large fraction of the lower corona.
They calculated orbits of fountain clouds ejected from regions in the Milky Way's spiral arms and run an MCMC minimization between the predicted $\hi$ emission from these clouds and the LAB data of complex C.
The results are shown in Fig.\ \ref{fig:complexC} and Table \ref{tab:evidence}.

The most massive and metal poor of all Galactic HVCs appears to be the result of a superbubble blowout in the Cygnus-Outer arm occurred 150 Myr ago.
The ejection of $\sim 3\times10^6 \mo$ of fountain gas has triggered the subsequent condensation of a similar amount of coronal material.
This lowered the rotation velocity and the metallicity of the fountain material.
The model is fitted only to the $\hi$ emission but it reproduces also the distance of complex C and its metallicity (see Fig.\ \ref{fig:complexC}, right).
The event that formed complex C requires an energy of $E\simeq 2\times 10^{54} \erg$, which can be achieved by $\sim 1\times 10^4$ supernovae assuming $20\%$ efficiency.
This is probably a rare event although ejections of similar $\hi$ masses are observed in spiral galaxies \citep{Boomsma+2008}. 
Most superbubble blowouts likely produce material at lower anomalous velocities that end up being observed as IVCs.
However, some (perhaps all) HVCs can be produced in this way.
The recent discovery of a fountain-like metallicity for the Smith Cloud indeed suggests a similar formation mechanism \citep{Fox+2016,Marasco&Fraternali2016}.

\section{Galaxy evolution with fountain accretion}
\label{sec:evolution}

In \S \ref{sec:observations} we have seen a number of successful comparisons between the predictions of the fountain accretion model and a variety of datasets.
In summary, the model reproduces the kinematics of the cold/warm gas at the interface of the galactic corona in great detail \citep{Marasco+2012,Fraternali+2013} and predicts the lag of the corona with respect to the disc recently measured in the Milky Way \citep{Marinacci+2011,Hodges-Kluck+2016}.
Most important of all, it accounts for accretion of low-metallicity material at a rate that is comparable to the SFR, providing an elegant solution to the gas accretion problem in local spiral galaxies.
In the future, much should be done to test the model against other observables and incorporate it into a self-consistent theory of galaxy evolution.
In this Section, we briefly describe two recent investigations in this direction that hint at the role of fountain accretion in the onset of metallicity gradients and the quenching of star formation.

\begin{figure}[ht]
\begin{center}
\sidecaption
\includegraphics[trim=0cm 0cm 0cm 0cm, clip=true, angle=-90, width=\textwidth]{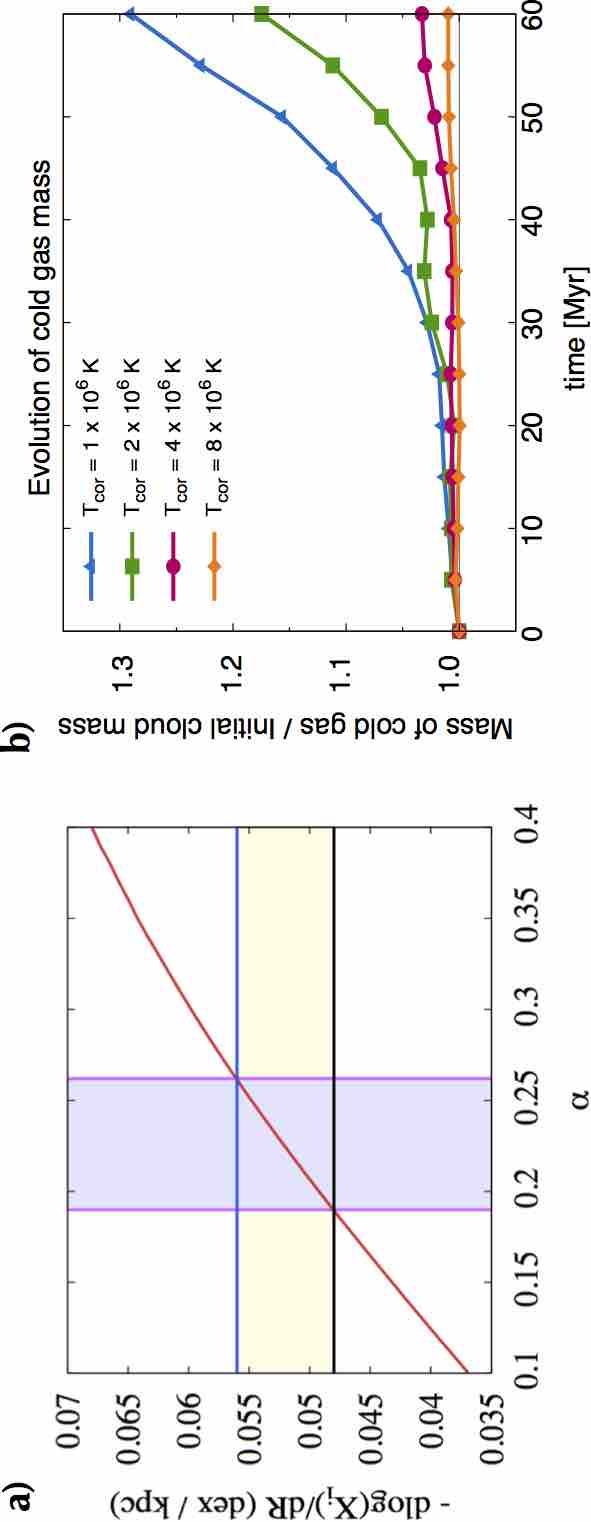}
\caption{
\emph{a) }
Abundance gradient (red curve) predicted for the Milky Way's disc as a function of the average rotation-velocity mismatch between the accreting and the disc gas ($\alpha=1-\frac{v_{\rm acc}}{v_{\rm disc}}$).
The shaded areas show the regions allowed by the data \citep{Luck&Lambert2011}, which correspond to material accreting with an average rotational velocity $v_{\rm \phi,\, acc}\simeq 170-195\kms$ in agreement with recent observations \citep{Hodges-Kluck+2016}; figure from \cite{Pezzulli&Fraternali2016}.
\emph{b) }
Corona condensation in the wakes of fountain clouds at the disc-halo interface as a function of the temperature of the corona.
The condensation is very efficient for temperatures similar or lower than that of the Milky Way ($T\simeq 2\times 10^6\K$) and it vanishes for temperatures above $4\times 10^6 \K$; figure from \cite{Armillotta+2016}.
}
\label{fig:newResults}
\end{center}
\end{figure}

\citet{Pezzulli&Fraternali2016} built a model of disc evolution in the presence of a generic gas accretion (not necessarily fountain-driven) characterized by a rotation velocity different from the local circular velocity.
In this scenario, which is very likely if the accreting gas comes from the condensation of the corona (\S \ref{sec:fountainAccretion}), the accreted material lands in the disc with a deficit of specific angular momentum.
The natural consequence is to induce inward radial flows of gas and the strength of these flows is regulated by the angular momentum mismatch between disc and accreting gas. 
These flows may be of great importance to bring gas from the outer regions of galaxies into the inner star forming regions. 
Luckily, they have a strong impact on abundance gradients, providing an easily observable test of the theory.
General analytic solutions for this problem have been found both in the absence or presence of inside-out growth of the disc \citep{Pezzulli&Fraternali2016}.
The current abundance gradient is steeper the higher the angular momentum mismatch.
Figure \ref{fig:newResults}a (red curve) shows the predicted gradient as a function of the average rotation-velocity mismatch of the accreting material for the Milky Way.
The observed gradients are compatible with accretion at about $70-80 \%$ of the local circular speed, i.e.\ with a  $v_{\rm lag}\simeq 45-70 \kms$ with respect to the disc rotation.
This value is very similar to the rotation of the corona 
that is required to fit the kinematics of the extraplanar gas in the Milky Way (\S \ref{sec:observations}).
Thus, it appears that metallicity gradients as steep as those observed in local galaxies are compatible with a supernova-driven prescription for gas accretion.

In a second work, \citet{Armillotta+2016} investigated the efficiency of condensation as a function of the coronal temperature using simulations that include thermal conduction and are akin to those shown in Fig.\ \ref{fig:condensation}.
The main result is shown in the right panel of Fig.\ \ref{fig:newResults}, where we see the amount of condensation with respect to the initial mass of the cold gas as a function of time.
Coronae with temperatures similar or lower than that of the Milky Way ($T\simeq 2\times 10^6\K$) cool very efficiently but this trend stops for temperatures above $\sim4\times 10^6 \K$.
If we assume that the coronal temperature is of order of the virial temperature then we can relate it to the virial mass of the dark matter halo.
We conclude that for virial masses larger than the Milky Way fountain-driven condensation and cooling of the corona become less and less efficient and these galaxies will struggle to gather new gas from their environment.
Gas accretion stops altogether for masses $M \gtrsim 10^{13} \mo$.

These findings may have important implications for the quenching of star formation in galaxies.
As the virial mass of a galaxy grows because of dark matter assembly, its ability to cool the surrounding corona decreases and it may end up consuming all the available gas and thus quench.
Moreover, when the galaxy enters a dense environment (group or cluster), its extended corona is extremely vulnerable to stripping and the galaxy will naturally become surrounded by gas at much higher temperature.
In these conditions, corona condensation stops and the star formation is starved of fuel.
Thus, both mass and environmental quenching \citep{Knobel+2015} could be accommodated into the fountain accretion mechanism.

\section{Concluding remarks}
\label{sec:conclusions}

In this Chapter, we have presented the physical principles of the fountain-driven (or supernova-driven) accretion model and its successes in reproducing observational data for local galaxies.
This model has been developed using tailored hydrodynamical simulations of the disc-corona interface. 
Physically motivated prescriptions from simulations have been also incorporated into large-scale galactic fountain models.
The comparison with the data is done by building artificial observations (in particular $\hi$ pseudo-datacubes) and minimizing the residuals between these and the actual data.
A general prediction is that, in the physical conditions typical of star forming galaxies, significant accretion occurs from the cooling of the lower corona at the disc-halo interface.
The cooling (condensation) is regulated by supernova feedback which can stimulate gas accretion at rates comparable with the galaxy SFR.

Cosmological simulations predict that galaxies with current virial masses above $10^{12} \mo$, after an initial phase of \emph{cold-mode} (filamentary) accretion \citep{Dekel+2009}, which lasts until $z\simeq 1-2$, move to a so-called \emph{hot-mode} accretion phase \citep{Keres+2009, Nelson+2013}. 
This accretion builds and feeds extended galactic coronae, which are unable to cool efficiently except at their very center where they can trigger episodic feedback from the central black hole \citep{Binney&Fraternali2012}.
Paradoxically, it is after $z\simeq 1-2$ that a galaxy like the Milky Way needs accretion of cold gas as the majority of its stars (at least the stars of the thin disk) are formed after $z=1$ \citep{Snaith+2015}.
Fountain-driven accretion provides an explanation for this apparent contradiction, as it triggers continuous cooling of the lower corona across the whole star-forming disc.
This mechanism could be the way star-forming galaxies extend their lives beyond the gas depletion time in the long epoch of hot-mode accretion.

We can also speculate that the shift of galaxies from the blue cloud to the red sequence is a consequence of a drop in the efficiency in gathering new gas.
This can happen because the corona becomes too hot, the galaxy falls into a larger halo (\S \ref{sec:evolution}) or because it simply loses its gaseous disc, e.g.\ due to mergers or gas stripping.
In the fountain accretion scenario, the key ingredient for efficient accretion is the presence of a star-forming disc of cold gas, which acts as the ``refrigerator'' of the corona. 
If a galaxy loses this disc it may take long time to reform it (or it may never happen) and it will permanently become a red early-type galaxy.

\begin{acknowledgement}
I thank Lucia Armillotta, James Binney, Antonino Marasco, Federico Marinacci and Gabriele Pezzulli for comments on the manuscript and Cecilia Bacchini for providing Figures 3 and 8. 
\end{acknowledgement}



\end{document}